\documentclass[fleqn,usenatbib]{mnras}

\usepackage{newtxtext,newtxmath}

\usepackage[T1]{fontenc}

\DeclareRobustCommand{\VAN}[3]{#2}
\let\VANthebibliography\thebibliography
\def\thebibliography{\DeclareRobustCommand{\VAN}[3]{##3}\VANthebibliography}

\usepackage{graphicx}	
\usepackage{amsmath}	
\usepackage{booktabs}
\usepackage{gensymb}
\usepackage{tablefootnote}
\usepackage{threeparttable}
\usepackage{subcaption}

\newcommand{\nickel}{$^{56}$Ni}
\newcommand{\mni}{M_\mathrm{Ni56}}

\usepackage[normalem]{ulem}

\title[Bolometric luminosity-width correlation]{All known Type Ia supernovae models fail to reproduce the observed bolometric luminosity-width correlation}

\author[Sharon \& Kushnir]{
	Amir Sharon$^{1}$\thanks{E-mail: amir.sharon@weizmann.ac.il},
	Doron Kushnir$^{1}$,
        and Nahliel Wygoda$^{2}$
	\\
	$^{1}$Dept.of Particle Phys. \& Astrophys., Weizmann Institute of Science, Rehovot 76100, Israel\\
	$^{2}$Department of Physics, NRCN, Beer-Sheva 84190, Israel\\
}

\date{Accepted XXX. Received YYY; in original form ZZZ}

\pubyear{2024}

\begin{document}
\label{firstpage}
\pagerange{\pageref{firstpage}--\pageref{lastpage}}
\maketitle

\begin{abstract}

Type Ia supernovae (SNe Ia) are widely believed to arise from thermonuclear explosions of white dwarfs (WDs). However, ongoing debate surrounds their progenitor systems and the mechanisms triggering these explosions. Recently, Sharon \& Kushnir showed that existing models do not reproduce the observed positive correlation between the $\gamma$-ray escape time, $t_0$, and the synthesized $^{56}$Ni mass, $\mni$. Their analysis, while avoiding complex radiation transfer (RT) calculations, did not account for the viewing-angle dependence of the derived $t_0$ and $M_\mathrm{Ni56}$ in multi-dimensional (multi-D) models during pre-nebular phases, where most observations performed. Here, we aim to identify an observational width--luminosity relation, similar to the $t_0$--$M_\mathrm{Ni56}$ relation to constrain multi-D models during pre-nebular phases while minimizing RT calculation uncertainties. We show that the bolometric luminosity at $t\le30$ days since explosion can be accurately computed without non-thermal ionization considerations, which are computationally expensive and uncertain. We find that the ratio of the bolometric luminosity at 30 days since explosion to the peak luminosity, $L_{30}/Lp$, correlates strongly with $t_0$. Using a sample of well-observed SNe Ia, we show that this parameter tightly correlates with the peak luminosity, $L_p$. We compare the observed $L_{30}/Lp$--$L_p$ distribution with models from the literature, including non-spherical models consisting of head-on WD collisions and off-centered ignitions of sub-Chandrasekhar mass WDs. We find that all known SNe Ia models fail to reproduce the observed bolometric luminosity-width correlation. 
\end{abstract}




\section{Introduction}
\label{sec:introduction}

Type Ia supernovae (SNe Ia) are widely accepted to be the result of thermonuclear explosions of white dwarfs (WDs), but their progenitor systems and explosion mechanism remain under debate \citep[for a review, see, e.g.,][]{Maoz2014}. Several theoretical scenarios for the progenitor systems have been suggested, including a Chandrasekhar mass (Chandra) or sub-Chandrasekhar mass (sub-Chandra) WD that ignites due to some external interaction and direct WD collisions. 

Predictions from these models have demonstrated agreement with certain observational characteristics of SNe Ia in specific instances. One particular set of characteristics frequently employed for model-observation comparisons is the Phillips relation \citep{Phillips1993,Phillips1999}, which relates the maximum flux with the width of the light curve in a specific wavelength band. This relation holds significance in cosmology, and its parameters are relatively straightforward to derive from observations, rendering it a widely adopted method for such comparisons \citep{Kasen2006,Sim2010,Blondin2017,Shen2021NLTE,Collins2022}. However, deriving these parameters from theoretical models and simulations is quite challenging. One of the key challenges in the calculation of these parameters and of most observable quantities is radiative-transfer (RT) calculations, which often prohibit a robust comparison to observations \citep[see reviews, e.g.][]{Hillebrandt2000,Noebauer2019}.

While advances in computational capabilities and theoretical understanding have significantly improved RT simulations, the large number of physical processes involved, particularly opacity from thousands of atomic transitions with both absorptive and scattering characteristics, prohibit comprehensive 3D calculations based on first principles. Various physical approximations — in particular, different treatments of the significant deviations from local thermodynamic equilibrium (LTE) — are employed by different RT codes to calculate the properties of the plasma and the radiation field. Additionally, approximate treatment of atomic physics is required due to the partially calibrated atomic data.

Recently, \cite{Blondin2022} compared various radiative transfer (RT) codes applied to identical benchmark models (the StaNdaRT public electronic repository\footnote{\url{https://github.com/sn-rad-trans}}). Among the codes considered, only CMFGEN \citep{Hillier1998, Dessart2010} performs full non-local thermodynamic equilibrium (NLTE) calculations from the onset of the explosion. The properties obtained from CMFGEN, such as light curves, temperature, and ionization profiles, behave differently than those from LTE codes after a few tens of days post-explosion, highlighting the importance of NLTE treatment during these phases. However, CMFGEN is computationally expensive and thus limited to 1D profiles.

One approach to bypass the radiative transfer (RT) calculation challenges is determining the $\gamma$-ray escape time, $t_0$, defined by \citep{Jeffery1999}
\begin{equation}
    f_\textrm{dep}(t) = \frac{t_0^2}{t^2},\:\:\:f_\textrm{dep}\ll 1,
\end{equation}
where $t$ is the time since explosion and $f_\textrm{dep}$ is the $\gamma$-ray deposition fraction, representing the portion of the generated $\gamma$-ray energy that is deposited in the ejecta.
This quantity can be estimated directly from the ejecta or using simple $\gamma$-ray radiation transfer simulations \citep{Stritzinger2006,Scalzo2014,Wygoda2019,Guttman2024}. To measure $t_0$ in observations of SNe, the ejecta must be sufficiently optically thin so that the bolometric luminosity, $L(t)$, equals the instantaneous deposited energy from radioactive decay:
\begin{equation}
\label{eq:L_eq_Q}
  L(t)=Q_\text{dep}(t).  
\end{equation}
For SNe Ia, this condition is typically met at $t\gtrsim60$ days post-explosion for most SNe \citep{Wygoda2019, Sharon2020}. The $\gamma$-ray escape time and the synthesized \nickel\ mass have recently been used to compare models with observations for various types of SNe \citep{Sharon2020, Sharon2020b, Sharon2023}. In \cite{Sharon2020b}, the $t_0$--$M_\mathrm{Ni56}$ distribution of SNe Ia was compared with model predictions from the literature. It was found that none of the models reproduced the observed relation, which showed a strong, positive correlation between the two parameters. This positive correlation, similar to the Phillips relation, implies that bright SNe are characterized by slowly evolving light curves and vice versa. Additionally, there is some scatter in the observed $t_0$--$M_\mathrm{Ni56}$ relation, which could be attributed to multi-dimensional (multi-D) effects from non-spherical explosions.

In multi-D models, the luminosity can be unevenly distributed across different viewing angles, even when the total luminosity, integrated over all angles, satisfies Equation~\eqref{eq:L_eq_Q}. This uneven distribution might cause the inferred $t_0$ and $M_\mathrm{Ni56}$ values to vary between viewing angles, deviating from their true values. Only at very late times ($t\gtrsim$150 day for typical SNe Ia), when the ejecta is in the deep nebular phase and sufficiently optically thin, does the emitted radiation become completely isotropic regardless of the ejecta’s structure. However, obtaining the bolometric luminosity at these late times is quite challenging because the supernova has significantly faded, and a large portion of the flux shifts to the near- and mid-infrared wavelengths. Only a few supernovae have the required temporal and wavelength coverage, such as the recent SN 2021aefx, having JWST observations at $t > 200$ days post-explosion \citep{Kwok2023,Chen2023}. Considering multi-D effects could explain the observed variability in the $t_0$--$M_\mathrm{Ni56}$ distribution, allowing for more stringent constraints on non-spherical models. However, radiative transfer (RT) calculations are necessary for this purpose, and, as previously mentioned, these calculations contain significant uncertainties.

One might argue that to address RT uncertainties and calculate $t_0$ for multi-D models, it is possible to extract the escape time from the calculated bolometric light curve by solely using times when Equation~\eqref{eq:L_eq_Q} holds. This method was applied to 1D ejecta in \cite{Kushnir2020}, and the resulting $t_0$ values were consistent with those obtained directly from the ejecta or through $\gamma$-ray RT simulations. However, a multi-dimensional structure within the ejecta complicates the use of this method to mitigate inaccuracies in bolometric light curve calculations performed by RT codes. The challenge arises because, although Equation~\eqref{eq:L_eq_Q} might hold for the angle-averaged luminosity at certain times, the distribution of luminosity among different observed angles remains unknown and cannot be verified by the $\gamma$-ray deposition. Therefore, the luminosity of a non-spherical ejecta at a given observed angle cannot be determined free of the uncertainties associated with RT.

In this paper, we seek an observational "width--luminosity" relation, similar to the $t_0$--$M_\mathrm{Ni56}$ relation, that would allow to constrain multi-D models at pre-nebular phases while minimizing the inherent uncertainties of RT calculations. We derive a relation that accomplishes this by imposing two constraints on the observational data. The first is that only the bolometric light curve is considered instead of observing specific bands. The second constraint is that the analysis is limited to early times, $t<30$ days since the explosion. We discuss the motivation for these constraints in Section~\ref{sec:shape_parameter}. 

Previous studies have examined the early-time bolometric properties of SNe Ia and explosion models \citep{Contardo2000,Stritzinger2006,Phillips2006,Scalzo2014,Scalzo2019,Wygoda2019,Gronow2021,Shen2021Multi}. Notably, the decline in bolometric magnitude during the first 15 days after the peak, $\Delta M_\text{bol}(15)$, was found to correlate well with the $\gamma$-ray escape time \citep{Stritzinger2006} and the peak luminosity, $L_p$ \citep{Scalzo2019}, although this relation was not as tight as in individual bands (i.e., the Phillips relation). The bolometric magnitude decline rate was also used in \cite{Shen2021Multi} and \cite{Gronow2021} to compare simulations of non-spherical sub-Chandra explosions with SNe Ia observations from \cite{Scalzo2019}. Their models broadly agreed with observations but failed to reproduce the bright segment of the observed relations, and the variations due to viewing angle were more pronounced than observed.




Our study presents several novel aspects:
(1) We select the luminosity ratio 30 days post-explosion to the peak luminosity, $L_{30}/L_p$, as our shape parameter. This parameter, which is independent of the supernova's brightness and distance, offers better model-constraining power and corresponds to earlier times when RT simulation results are more reliable, compared to the commonly used $\Delta M_\text{bol}(15)$; (2) We analyze several different SNe Ia explosion models from the literature, including 1D models of Chandra and sub-Chandra explosions and non-spherical 2D models of head-on collisions and off-centered sub-Chandra ignitions; (3) We utilize a high-quality sample of well-observed SNe Ia from \cite{Sharon2020} and supplement it with additional objects. By focusing on early-time observables, we require less stringent observational requirements than those needed for determining $t_0$ and $\mni$, necessitating continuous observations for at least ${\approx}100$ days post-explosion. As a result, early-time observables can be measured for a larger fraction of supernovae, significantly increasing the sample size.




The paper is organized as follows: In Section~\ref{sec:data}, we describe our observational data, the sample of SNe Ia, and the explosion models we have incorporated. Section~\ref{sec:shape_parameter} outlines our motivation for using the early bolometric light curve to constrain models. The resulting $L_{30}/L_p$--$L_p$ relation is presented in Section~\ref{sec:results} and illustrated in Figure~\ref{fig:Lp_LoverLp}. In Section~\ref{sec:opacity}, we examine the opacity of the 1D models to account for the short rise times of the low-luminosity models. Our findings are discussed and summarized in Section~\ref{sec:discussion}. Appendix~\ref{app:NT_ionization} estimates when the ionization levels of SNe Ia ejecta depart from LTE. Appendix~\ref{app:opacity} compares the rise time of our models to a simplified analytical solution \citep{KushnirKatz2019} to evaluate the effects of the ejecta opacity. In Appendix~\ref{app:log_graphs}, we provide some of our results in magnitude space for completeness. Tables containing the parameters of both the observed sample and the models are provided in Appendix~\ref{app:parameters}.


\section{Observed data and models}
\label{sec:data}

\subsection{Observations}
\label{sec:observations}

The observational data comprises the SNe Ia sample from \citet{Sharon2020} and \cite{Sharon2023}, with some modifications\footnote{Several SNe were excluded upon reevaluation due to incomplete light curve coverage at peak, and the distance to SN 2011fe was updated using the latest Cepheid-based distance from \citet{Riess2022}.}, augmented with SNe lacking late-time observations required for measuring $t_0$ and $\mni$, yet possessing sufficient early-time data for this analysis. These additional SNe, sourced from the Carnegie Supernova Project \citep{Contreras2010,Stritzinger2011,Krisciunas2017,Burns2018}, expand the sample size to 47 SNe. The photometry, distance, extinction values, and explosion time are taken from the literature. All SNe in the sample have near-infrared (NIR) photometry, contributing ${\approx}20-40$\% of the total flux at times considered in this analysis \citep{Scalzo2014,Sharon2023}. Bolometric light curves are constructed using the methods described by \cite{Sharon2020}, and are available in the online supplementary material.

Following the approach of \cite{Sharon2020b}, we construct the $t_0$--$\mni$ distribution of the sample (only for objects with adequate observations) using the methods of \cite{Sharon2020}. The results, shown in Figure~\ref{fig:t0-Ni} as black symbols, align closely with the distribution in \cite{Sharon2020b}, as most of the samples overlap. This comparison is instructive for contrasting the bolometric width--luminosity relation~(Section \ref{sec:results}).

Utilizing the sample of bolometric light curves, the analysis parameters for each SN are calculated as follows. The peak luminosity and time are determined by fitting a low-order polynomial to all luminosity measurements $\{L_i\}_{i=1}^N$ with $L_i>0.85\cdot\max(L_i)$. The uncertainty in peak luminosity includes statistical errors and several systematic terms, such as uncertainties in distance, extinction, and missing UV flux (missing flux from longer wavelengths is negligible due to the required NIR coverage), all of which are propagated to the peak luminosity error. The value of $L_{30}$ is obtained by linear interpolation. The error in the $L_{30}/L_p$ parameter (see Section~\ref{sec:shape_parameter}) is unaffected by distance uncertainty and only minimally influenced by uncertainties in extinction and missing UV flux. However, the error includes a contribution from the unknown explosion time, estimated by assuming the explosion time uncertainty is 5\% of the time to the first measurement, $\Delta t_\mathrm{exp}=0.05(t_1-t_\mathrm{exp})$, where $t_1$ and $t_\mathrm{exp}$ are the times of the first luminosity measurement and the estimated explosion time, respectively. These errors are consistent with the SNe Ia rise time errors derived in \cite{Ganeshalingam2011}, which were within 1.5 days.

The derived parameters for the SNe sample are presented in Table~\ref{tab:sn}.

\subsection{Models}
\label{sec:models}

The models ejecta analyzed in this work include 1D and 2D models from the literature. The 1D models encompass explosions of Chandra WDs \citep{Dessart2014}, as well as centrally ignited sub-Chandrasekhar CO WDs with masses ranging from $0.8$ to $1.1\,M_\odot$ \citep{Kushnir2020}. The 2D models include head-on WD collisions \citep{Kushnir2013} and off-centered ignited sub-Chandra CO WDs \citep{Schinasi2024}. These 2D models are axisymmetric, with their luminosity depending on the observed inclination angle $\theta$, measured relative to the symmetry axis ($\theta=0,180\degree$ along the symmetry axis). The off-centered ignited sub-Chandra models have the same mass range as the 1D models and include a parameter $z_\text{ig}=0,0.1,0.2,0.5$, representing the ignition position along the symmetry axis normalized by the WD radius ($z_\text{ig}=0$ corresponds to central ignition). 

\begin{figure}
	\includegraphics[width=\columnwidth]{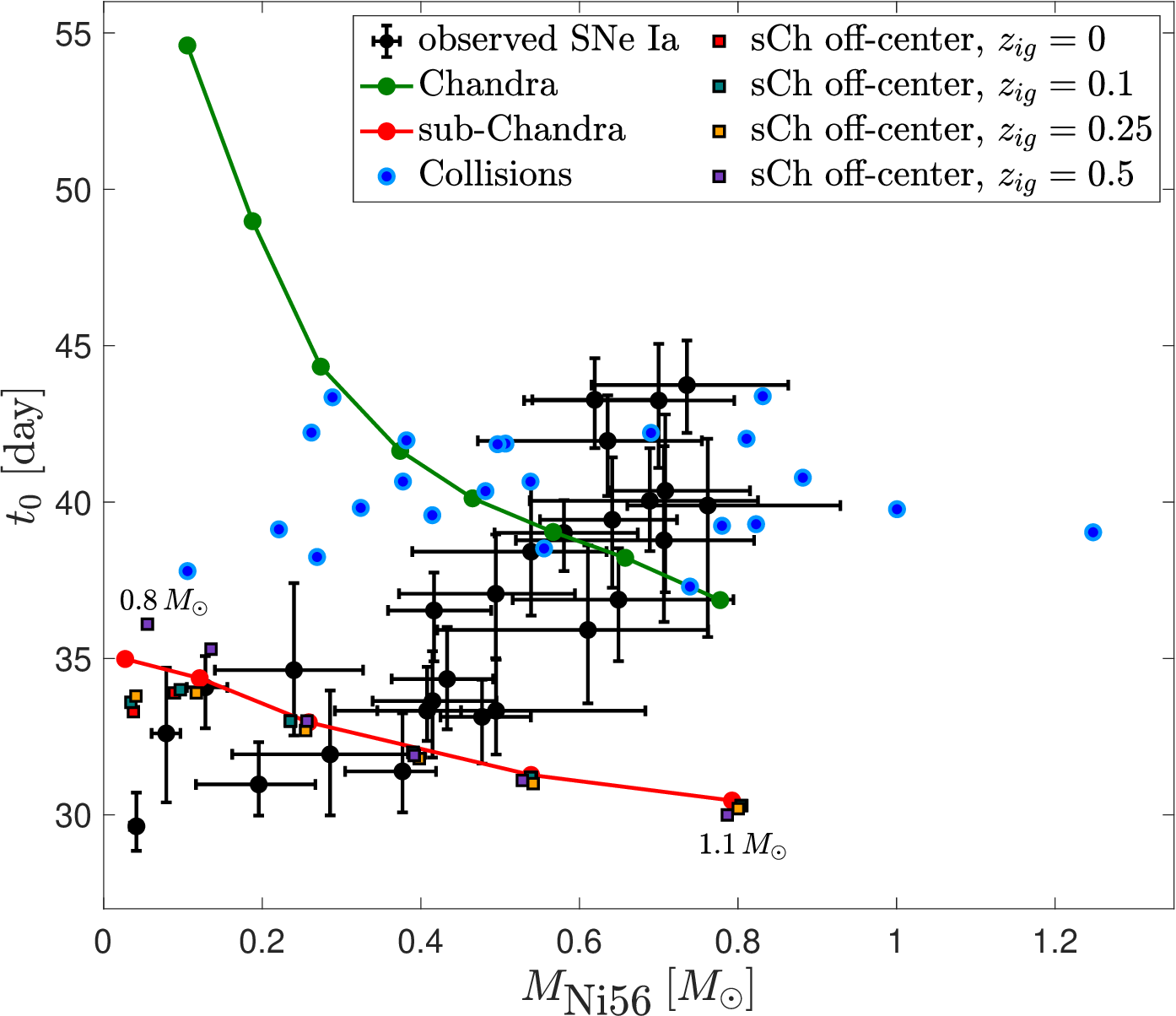}
    \caption{The distribution of the \nickel\ mass, $\mni$, and the $\gamma$-ray escape time, $t_0$, for the observed sample and explosion models. The symbols are the same as in Figure~\ref{fig:Lp_LoverLp}. The lowest and highest sub-Chandra progenitor mass is displayed near their respective values. Our results replicate the findings of \citet{Sharon2020b,Schinasi2024}. Note that the low-luminosity sub-Chandra $z_\textrm{ig}=0$ ignitions show slight deviations from their 1D counterparts due to 2D instabilities in the thermonuclear detonation wave \citep[see][for a detailed discussion]{Schinasi2024}. The conclusion that none of the models accurately reproduce the observed distribution remains unchanged.}
    \label{fig:t0-Ni}
\end{figure}

In all models, the explosion was simulated until the ejecta reached a state of free streaming, resulting in a homologously expanding profile. At this stage, the models' $\gamma$-ray escape time, $t_0$, can be computed through simple $\gamma$-ray transfer simulations or direct analytical calculations \citep{Wygoda2019}. For consistency with \cite{Sharon2020b}, we present the results using the analytical calculation in this work, noting that the differences compared to simulations are within a few percent. The $t_0$ results and the synthesized \nickel\ mass from the explosion are provided in Table~\ref{tab:models} and displayed in Figure~\ref{fig:t0-Ni}, replicating the findings of \cite{Sharon2020b,Schinasi2024}. Note that the low-luminosity sub-Chandra $z_\text{ig}=0$ ignitions show slight deviations from their 1D counterparts due to 2D instabilities in the thermonuclear detonation wave \citep[see][for a detailed discussion]{Schinasi2024}. The conclusion that none of the models accurately reproduce the observed distribution remains unchanged.

To determine the models' bolometric width--luminosity relation, we conducted RT simulations with the provided ejecta profiles to compute their bolometric light curves. These simulations were performed using the URILIGHT code \citep{Wygoda2019,Wygoda2019b}, a Monte Carlo code based on the approximations used in the SEDONA program \citep{Kasen2006Sedona}. The key assumptions include homologous expansion, an LTE configuration for the ions' ionization state and energy level occupation, and expansion opacities with optical depths calculated using the Sobolev approximation \citep[with the method of][]{PintoEastman2000}. The thermalization parameter $\epsilon$, which defines the probability that a photon absorbed in a given transition is re-emitted in a different transition, was set to 0.8.

\section{The bolometric light curve shape parameter}
\label{sec:shape_parameter}
This section presents the rationale for selecting the $L_{30}/L_p$ shape parameter as an observable that can robustly constrain models. In Section~\ref{sec:Identifying}, we identify the observed quantities RT simulations can produce with minimal uncertainty, limiting our analysis to these results. Next, we identify observables within these constraints to compare models and observations effectively (Section~\ref{sec:Selecting}).

\subsection{Identifying Minimally Uncertain Observed Quantities}
\label{sec:Identifying}

The first constraint we impose is limiting the analysis to the bolometric light curve rather than comparing specific bands. The reason is that the simulated flux of a specific band is significantly more sensitive to uncertainties in the opacities of bound-bound transitions. These uncertainties can arise from incomplete atomic data, the opacity calculation method, and the material's ionization state. The re-emission of absorbed photons redistributes the flux to other wavelengths, typically from the UV to the IR, making the total flux less susceptible to these uncertainties \citep{Gronow2021,Shen2021NLTE}. To illustrate this claim, we use the StaNdaRT public electronic repository, focusing on the toy06 benchmark model, a $1\,M_\odot$ model with characteristic Ia SNe \nickel\ mass ($0.6\,M_\odot$) and kinetic energy ($E_\text{kin} = 10^{51}\,\textrm{erg}$) values, and analytic density and composition profiles.

\begin{figure*}
	\includegraphics[width=0.9\textwidth]{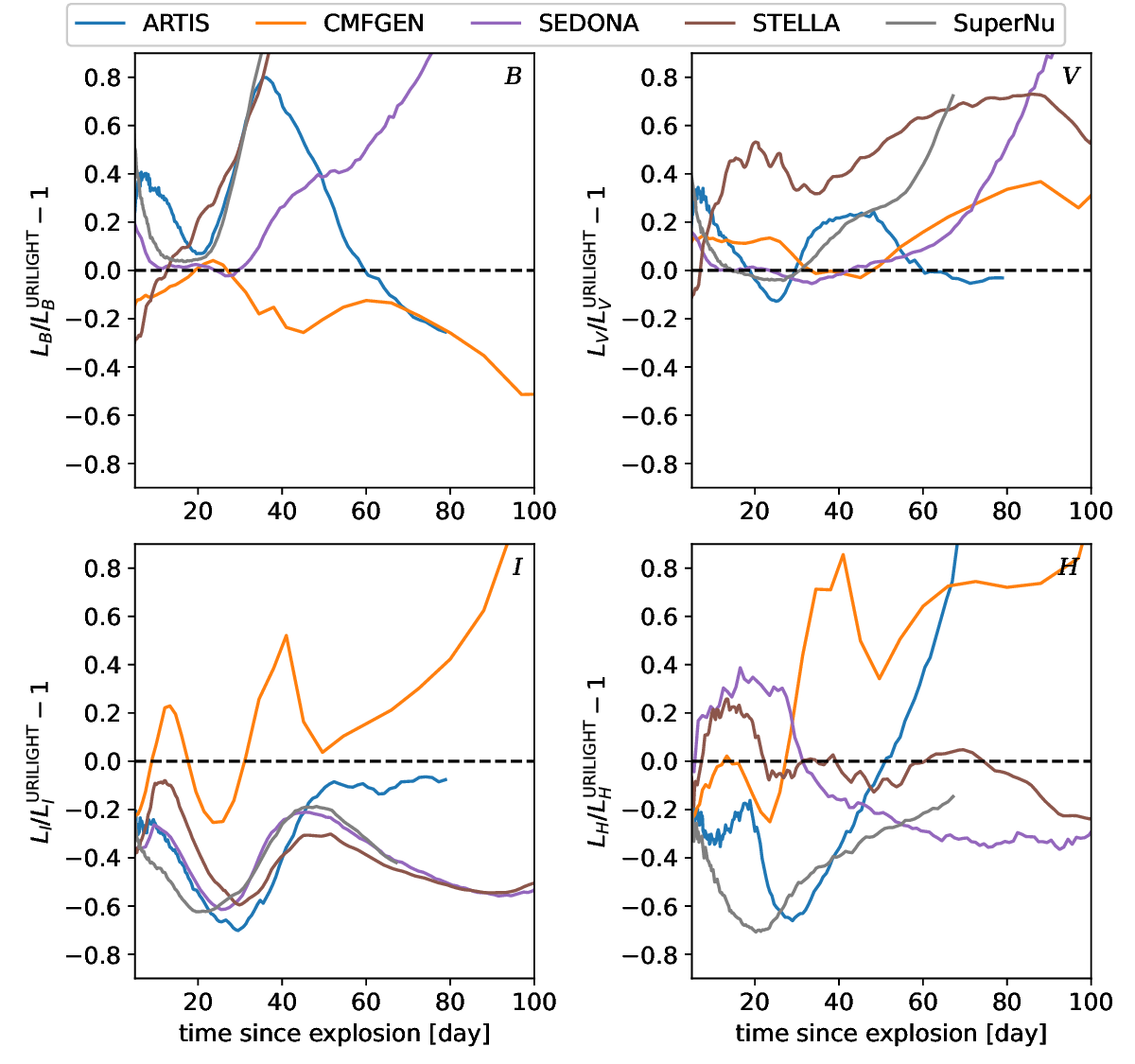}
    \caption{Comparison of multi-band light curves generated by various RT codes with respect to the URILIGHT code, performed on the toy06 model from \protect\cite{Blondin2022}. All data are taken from the StaNdaRT public electronic repository. Plotted is the code's deviation from URILIGHT, in flux space, for the $B$, $V$, $I$, and $H$ bands. All bands except the $V$ band show deviations of tens of percents. The $V$ band light curves are in better match, with deviations of up to $25$ percent for all the codes except STELLA. After 30 days, deviations in the $B$ and $H$ bands substantially increase.}
    \label{fig:mag_compare}
\end{figure*}

\begin{figure}
	\includegraphics[width=\columnwidth]{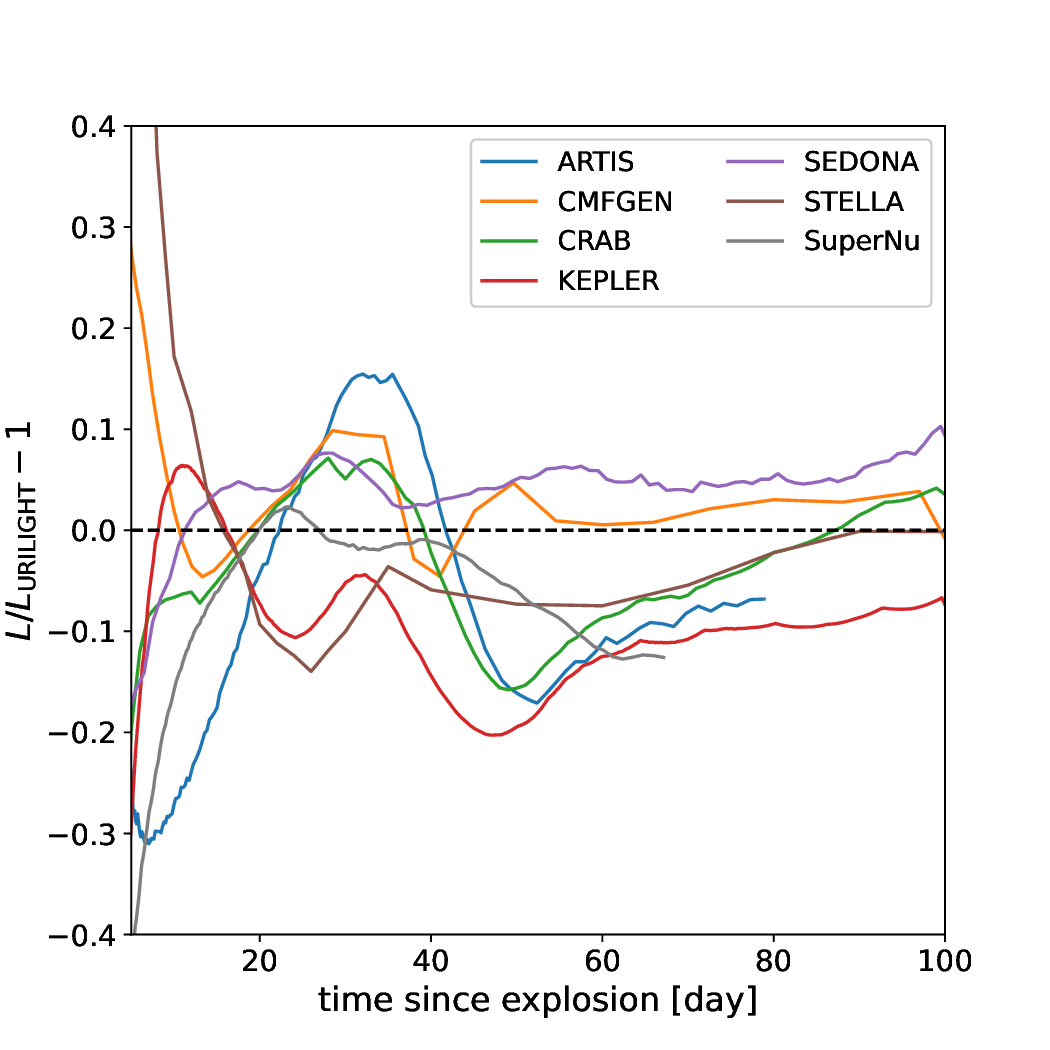}
    \caption{Same as Figure~\ref{fig:mag_compare} but for the bolometric luminosity, and the comparison is against a larger set of models. Deviations here are smaller than individual bands an, apart from the initial ${\approx}10$ days, are mostly within ${\approx}15$ percent.}
    \label{fig:lum_compare}
\end{figure}

In Figure~\ref{fig:mag_compare}, we compare the multi-band light curves (in flux space) of the toy06 model from various RT codes\footnote{The RT codes are: ARTIS \citep{Shingles2020}, CMFGEN \citep{Hillier1998}, CRAB \citep{Utrobin2004}, KEPLER \citep{Woosley2002}, SEDONA \citep{Kasen2006Sedona}, STELLA \citep{Blinnikov1993}, SUMO \citep{Jerkstrand2011PhD}, and SuperNu \citep{Wollaeger2013}.} against the results of the URILIGHT code. In most bands, differences are tens of percent at most times and can exceed 50 percent. Apart from the STELLA code, variations are less pronounced in the $V$ band but can still reach ${\approx}25$ percent. Generally, codes that show over-luminous light curves in one band tend to have under-luminous light curves in another band due to flux redistribution. Figure~\ref{fig:lum_compare} illustrates the same comparison for the bolometric luminosity. As shown in the figure, the deviations in bolometric luminosity are smaller than in individual bands and, apart from the initial ${\approx}10$ days, are mostly within ${\approx}15$ percent.

Another effect observable in Figure~\ref{fig:mag_compare} is that starting from $\approx30$ days post-explosion, the deviations significantly increase in certain bands. Additionally, the light curves of CMFGEN, the only NLTE code used during non-nebular times, deviate noticeably from the other codes. We, therefore, attribute this increased deviation at later times to the impact of NLTE effects, which become more pronounced as the density decreases. The deviation from LTE is driven by ionization and excitation caused by the thermalization of high-energy (up to a few MeV) non-thermal leptons produced in the \nickel\ decay chain, either as decay products or as electrons up-scattered by $\gamma$-rays from the decay \citep{Axelrod1980,Kozma1992,Baron1996}. This deviation significantly affects the material's opacity and the SNe light curves. Non-thermal ionization is known to be the dominant ionization process in the nebular phase and, as we demonstrate below, is also significant at earlier times.





\begin{figure}
	\includegraphics[width=\columnwidth]{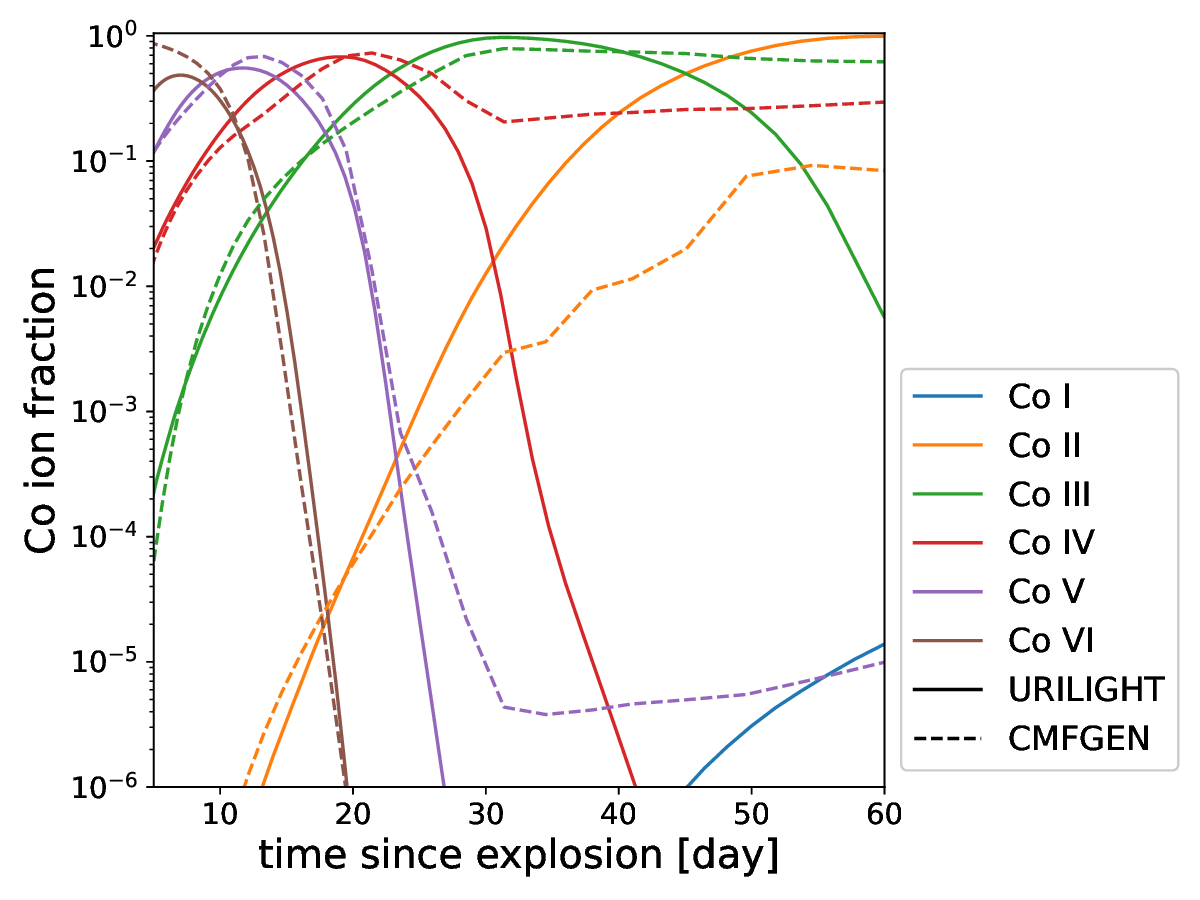}
	\includegraphics[width=\columnwidth]{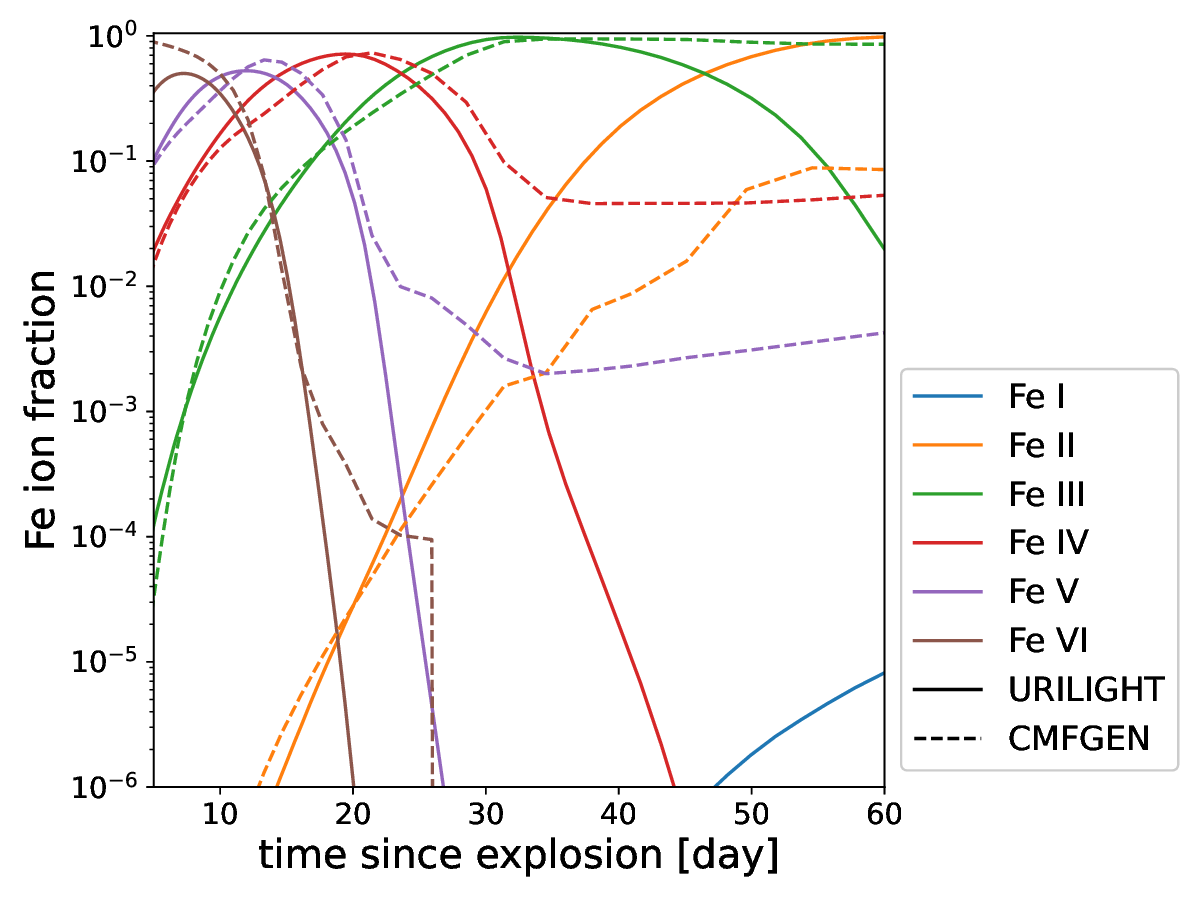}
    \caption{Temporal evolution of the mean Co (top) and Fe (bottom) ionization levels in the toy06 model, depicted for URILIGHT (solid lines) and CMGFEN (dashed lines). The ionization levels show good agreement up to ${\approx}30$ days from the explosion, after which they diverge. This is attributed to the epoch when ionization from non-thermal leptons becomes significant.}
    \label{fig:ion_compare}
\end{figure}

To evaluate the significance of NLTE ionization, we again utilize the StaNdaRT repository to compare the evolution of ionization structures between URILIGHT (LTE) and CMFGEN (NLTE). Figure~\ref{fig:ion_compare} shows the results for the toy06 model, focusing on the fractional ionization levels of Co and Fe averaged across the entire ejecta. Up to ${\approx}30$ days post-explosion, the ionization levels from both codes are in good agreement. However, after this period, the ionization levels begin to diverge, with the CMFGEN ejecta becoming significantly more ionized. This divergence marks the onset of significant ionization from non-thermal leptons. Note that the abundance of Co V is lower compared to Fe V in the CMFGEN simulation. This discrepancy arises because non-thermal ionization is not included for Co IV due to a lack of atomic data \footnote{John Hillier, private communication.}. This limitation further demonstrates the uncertainties inherent in these calculations.

In Appendix~\ref{app:NT_ionization}, we use simple analytical considerations to determine when non-thermal ionization becomes significant. Our calculations show that this epoch is around 30 days post-explosion, consistent with previous findings. Consequently, we limit our analysis to the early bolometric light curve, $t\le30$ days from the explosion, corresponding to times when the LTE assumption holds. Our approach aligns with the findings of \cite{Shen2021NLTE}, who compared light curves of sub-Chandrasekhar explosions using both LTE and NLTE RT codes. They observed that 15 days past peak light, significant discrepancies arise between the two methods in the observed magnitudes across most bands, while differences at peak light are much smaller. Moreover, the bolometric luminosities were roughly similar during the considered times, up to approximately 30 days since peak light. They also examined the effect of the absorption parameter, $\epsilon$, in LTE simulations and found that altering its value had minimal impact on the bolometric light curve, though individual magnitudes varied starting from 15 days after peak light.

\subsection{Selecting Observables for Model Comparison}
\label{sec:Selecting}

\begin{figure}
	\includegraphics[width=\columnwidth]{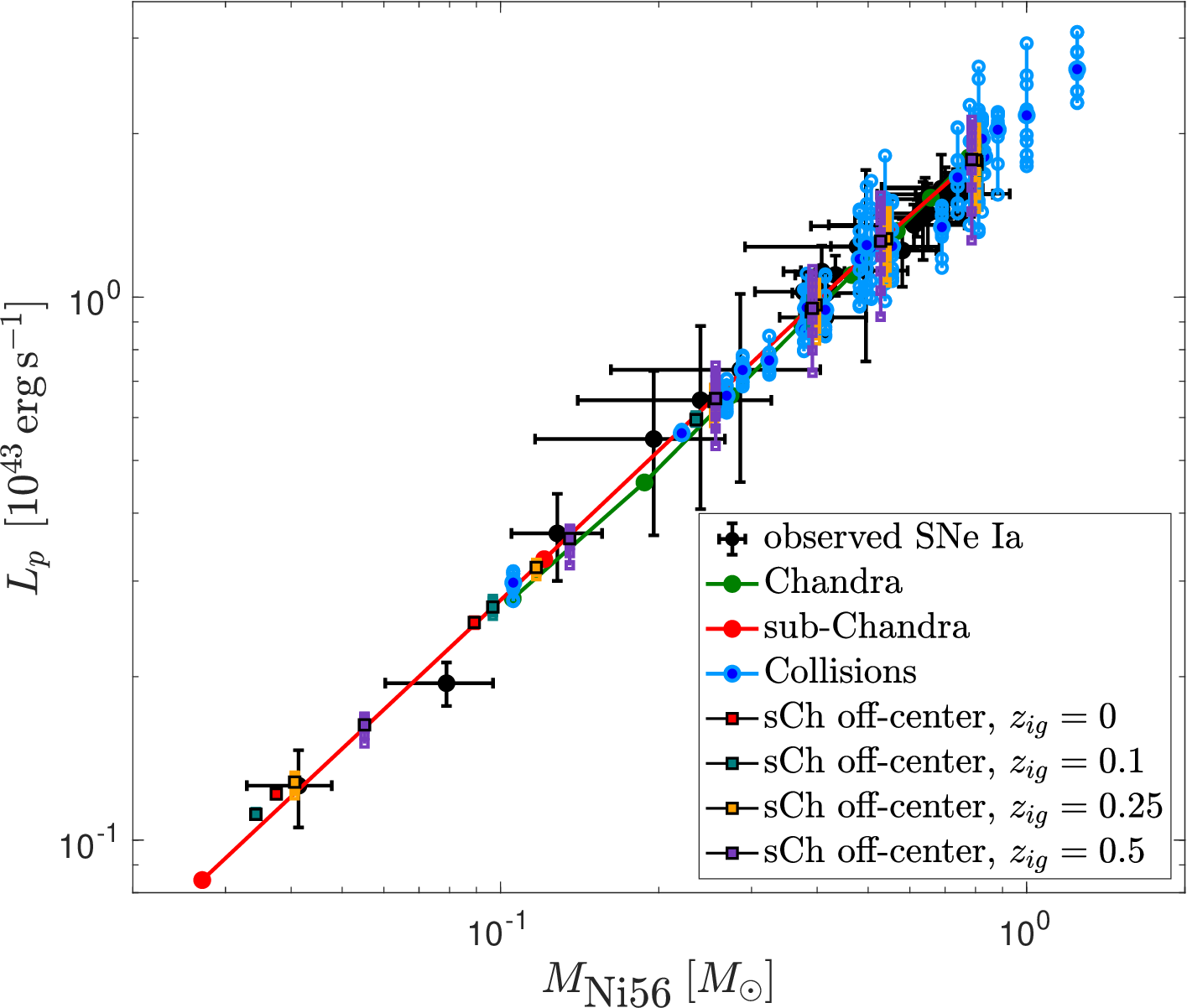}
    \caption{Comparison between the \nickel\ mass, $\mni$, to the peak luminosity, $L_p$, for the observed sample and explosion models. The symbols are the same as in Figure~\ref{fig:Lp_LoverLp}. The two parameters have an almost perfect correlation. }
    \label{fig:ni_lp}
\end{figure}

We aim to construct a width--luminosity relation under the constraints from above, similar to the Phillips relation and the $t_0\text{--}\mni$ distribution. The luminosity parameter we choose is, naturally, the peak bolometric luminosity. In Figure~\ref{fig:ni_lp} we compare $L_p$ with the late-time luminosity parameter, $\mni$. Consistent with previous works \citep[e.g.,][]{Scalzo2019}, these parameters for the observed SNe Ia, shown as black symbols, display an almost perfect correlation, with a Pearson correlation coefficient of $\rho=0.984$. 

\begin{figure}
	\includegraphics[width=\columnwidth]{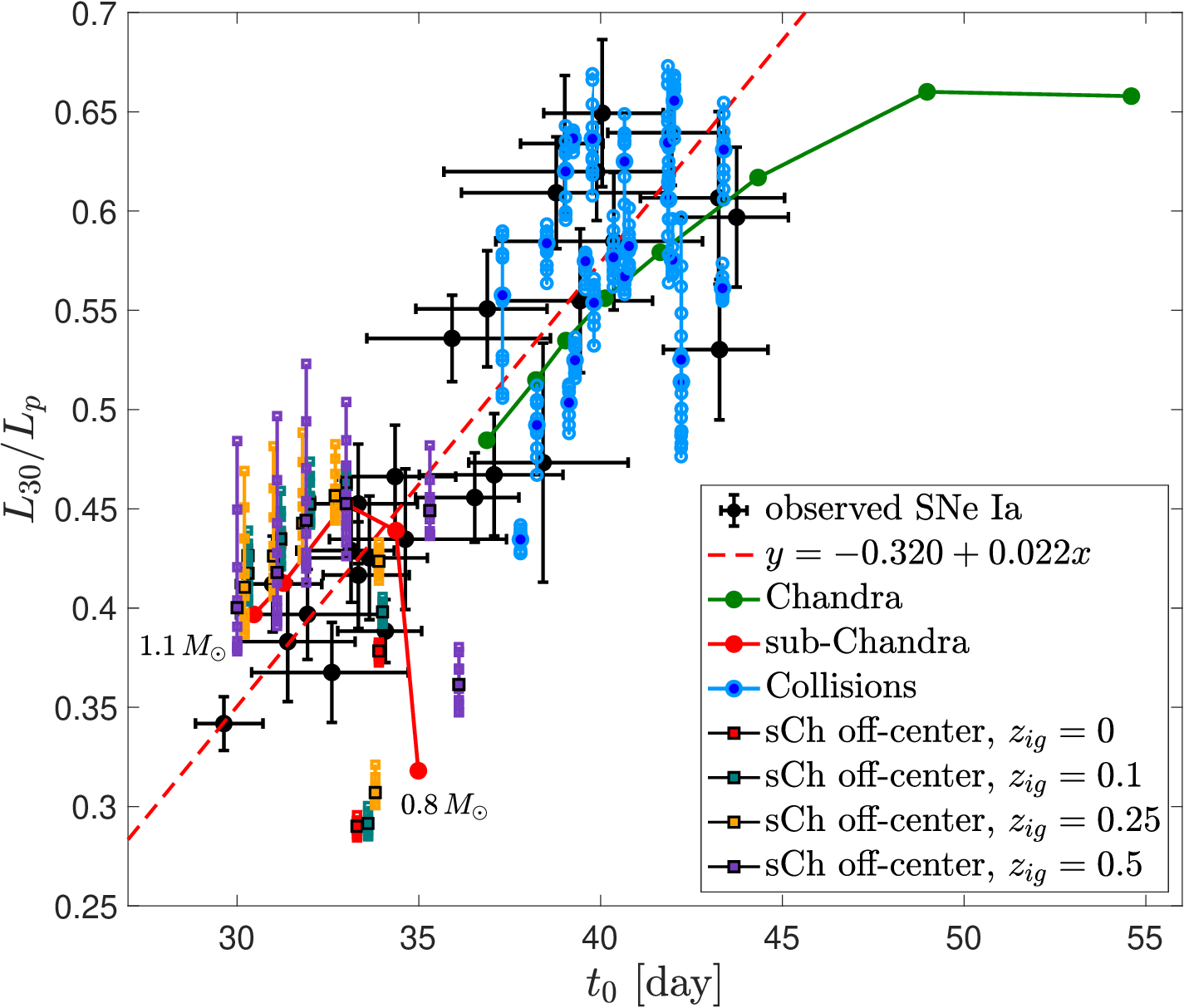}
    \caption{Same as Figure~\ref{fig:ni_lp} but for the $\gamma$-ray escape time, $t_0$, and $L_{30}/L_p$. For the sub-Chandra models, the lowest and highest progenitor mass is displayed near their respective values. The parameters exhibit strong correlations in both the observations and the models, except for the sub-Chandra detonations. These models show a decrease of $L_{30}/L_p$ for the high $t_0$ values, corresponding to low-mass WD progenitors. This drop is related to the short rise times of these models; see the main text for a detailed discussion. We also fit a linear regression model to the observed sample (dashed red line), and we find a slope of $\approx0.022\,\text{day}^{-1}$.}
    \label{fig:t0_LoverLp}
\end{figure}

The selection of the shape parameter is less trivial and we choose the luminosity ratio 30 days after the explosion to the peak luminosity, $L_{30}/L_p$. In Figure~\ref{fig:t0_LoverLp}, we compare $L_{30}/L_p$ with the late-time shape parameter, $t_0$. Here, the correlation of the observed sample is also prominent but to a lesser extent, with a Pearson correlation coefficient of $\rho=0.872$. We also fit a linear regression model between the two parameters, represented by $L_{30}/L_p=a\cdot t_0+b$. The results, shown by a dashed red line, have a slope of $a\approx0.022\,\text{day}^{-1}$. This approach aims to compare the ability to constrain models with the $L_{p+15}/L_p$ parameter (see below). Given the strong correlations evident in Figures~\ref{fig:ni_lp} and~\ref{fig:t0_LoverLp}, we expect the positive correlation observed in the $t_0$--$\mni$ relation to hold for the bolometric width--luminosity relation as well.

\begin{figure}
	\includegraphics[width=\columnwidth]{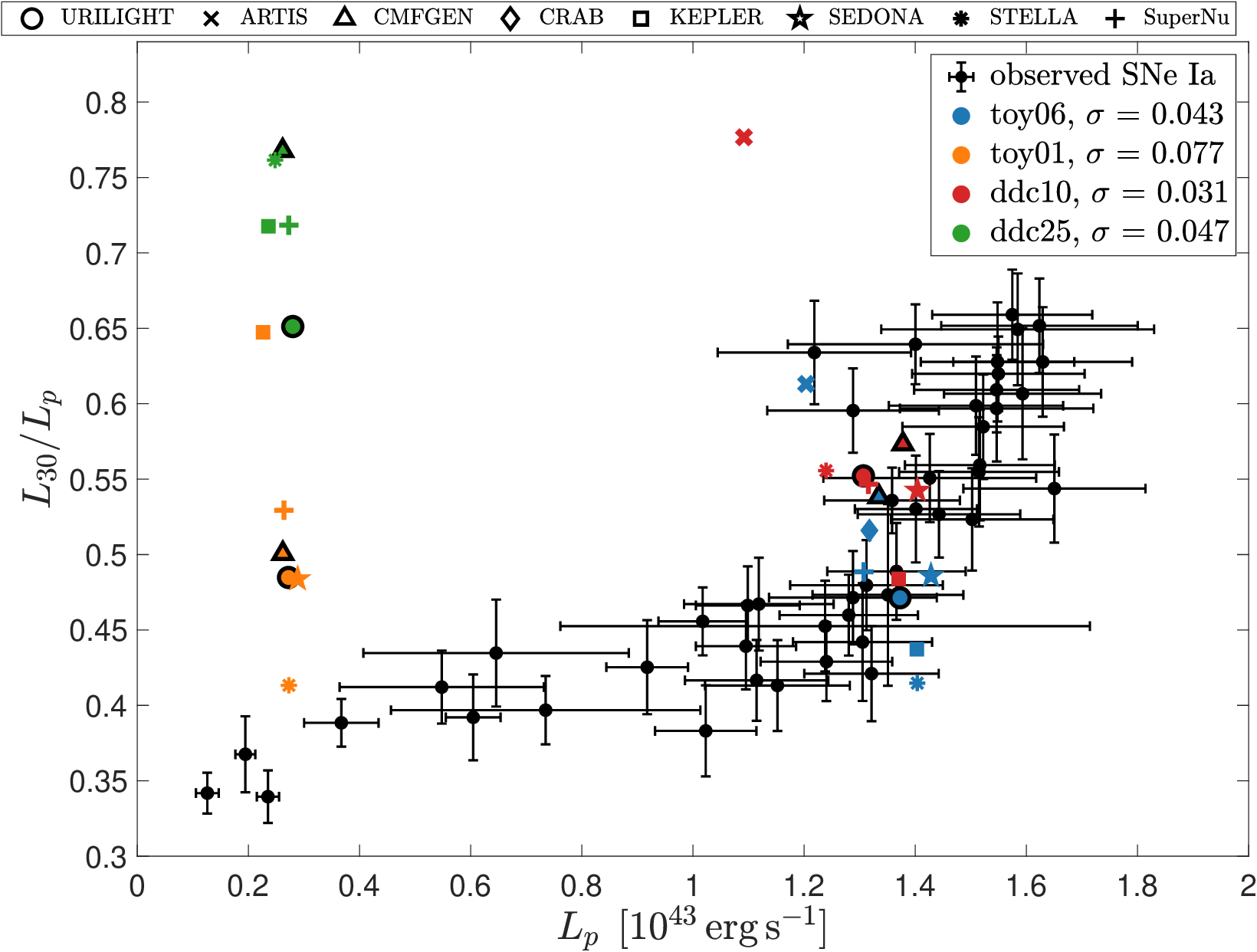}
    \caption{The $L_p$--$L_{30}/L_p$ relation of the observed sample (black symbols) and the models from the StaNdaRT repository. The StaNdaRT results are represented by the colored symbols, with the colors and shapes corresponding to different explosion models and RT codes, respectively. The scatter in the $L_{30}/L_p$ parameter between the RT codes for each explosion model (excluding ARTIS; see text for details) is provided in the legend.
    }
    \label{fig:Lp_L30overLp_standart}
\end{figure}

Figures~\ref{fig:ni_lp} and~\ref{fig:t0_LoverLp} also present the same comparisons for the explosion models. For non-spherical models, the $L_p$ and $L_{30}/L_p$ results are shown for different values of the inclination angle $\theta$, evenly distributed in $\cos{\theta}$, along with their angle-averaged values. A strong correlation is evident within the models' $\mni$--$L_p$ distribution, and they align with the observed relation (Figure \ref{fig:ni_lp}). However, the behavior of the shape parameters in Figure~\ref{fig:t0_LoverLp} is more complex. While the models generally follow the observed relation, the collision and Chandra models represent the slowly evolving SNe Ia, whereas the sub-Chandra models only cover the quickly evolving region (this is also evident in Figure~\ref{fig:t0-Ni}). Notably, the $t_0$--$L_{30}/L_p$ relation of the sub-Chandra models is non-monotonic. This occurs because low-mass WD progenitors exhibit short rise times (from explosion to peak light) despite having longer $\gamma$-ray escape times than higher-mass progenitors. We discuss this behavior further in Section~\ref{sec:opacity}.
 a $1\,M_\odot$ model with characteristic Ia SNe \nickel\ mass ($0.6\,M_\odot$) and kinetic energy ($E_\text{kin} = 10^{51}\,\textrm{erg}$)
Finally, we compare the $L_p$--$L_{30}/L_p$ relation of the models and RT codes in the StaNdaRT repository. Figure~\ref{fig:Lp_L30overLp_standart} displays the observed SNe Ia relation along with the RT code results for the four benchmark ejecta models in the StaNdaRT repository. These models include, in addition to the toy06 model described above, the toy01 model, which is identical to the toy06 model except for a lower \nickel\ yield of $0.1,M_\odot$. They also include two models from the Chandra WD explosions of \cite{Dessart2014}, the ddc10 and ddc25 models, with \nickel\ yields of $0.12$ and $0.52,M_\odot$, respectively (note that not all of the RT results are available for all models). We calculate the scatter in the $L_{30}/L_p$ parameter for each explosion model among the RT codes, excluding the ARTIS results due to their significant deviation from the others, which is most likely a result of the long rise times of these models. The scatter for the 'normal' luminosity models, toy06 and ddc10, is ${\approx}9$ and ${\approx}6$ percent, respectively. For the low-luminosity models, toy01 and ddc25, the scatter is ${\approx}15$ and ${\approx}6$ percent, respectively. The small scatter in $L_{30}/L_p$ between models supports our choice for the shape parameter. 



\begin{figure}
    \includegraphics[width=\columnwidth]{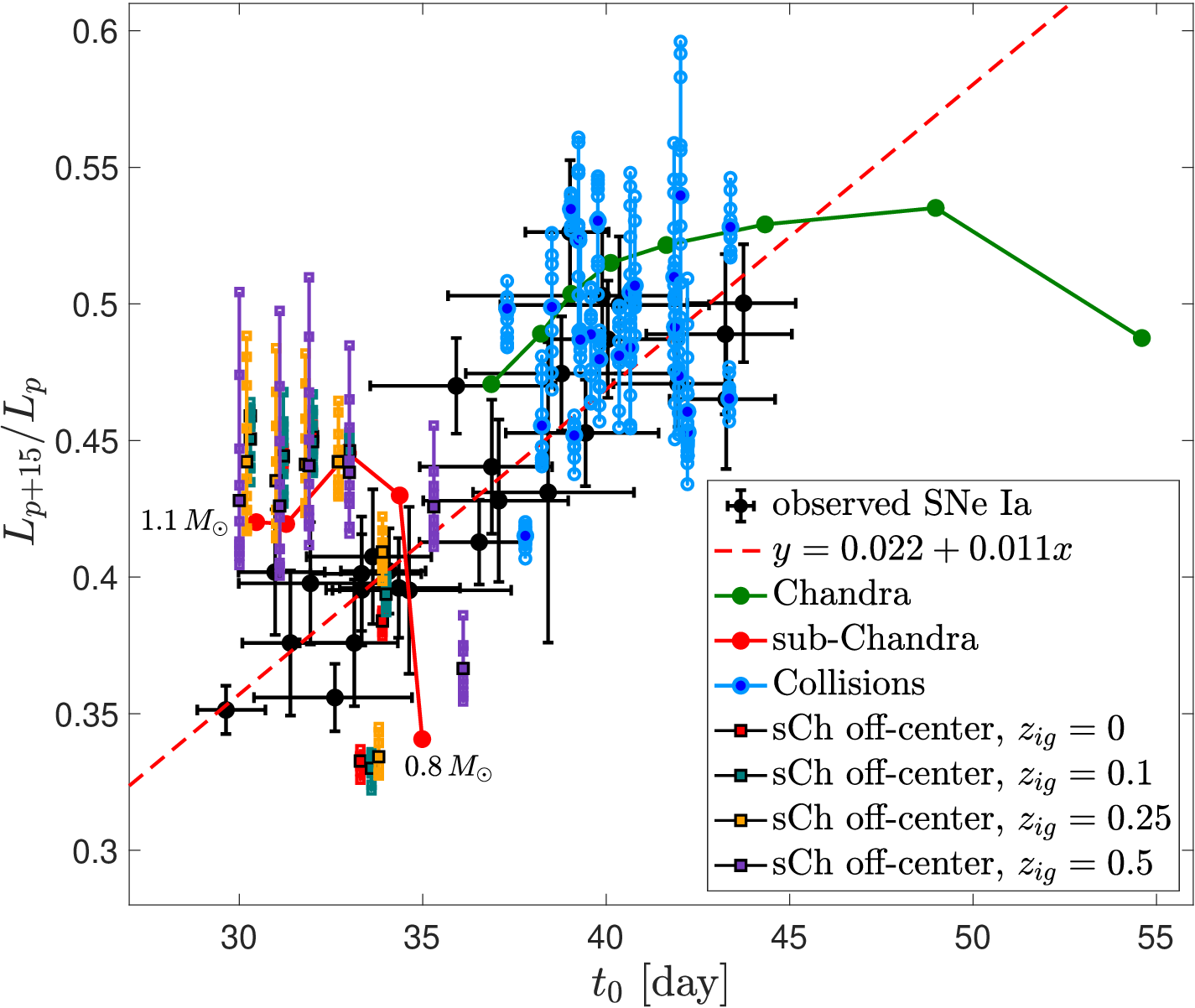}
    \caption{Same as Figure~\ref{fig:t0_LoverLp} but for the $L_{p+15}/L_p$--$t_0$ relation. A version of this figure in magnitude space is provided in Appendix~\ref{app:log_graphs}.}
    \label{fig:t0_Lp15overLp}
\end{figure}

We also explore replacing $L_{30}$ with the luminosity 15 days post-peak, $L_{p+15}$, similar to the bolometric magnitude decline rate, $\Delta M_\text{bol}(15)=-2.5\log(L_{p+15}/L_p)$, used in previous studies \citep{Stritzinger2006,Scalzo2019,Shen2021Multi}. In \cite{Stritzinger2006}, for a sample of 16 SNe Ia, this parameter was found to correlate well with $t_0$. In \cite{Scalzo2019}, it was shown to correlate with $\mni$ and $L_p$. Figure~\ref{fig:t0_Lp15overLp} presents the $L_{p+15}/L_p$--$t_0$ distribution of our sample. We use $L_{p+15}/L_p$ instead of the more commonly used form of $\Delta M_\text{bol}(15)$ to facilitate better comparison with the $L_{30}/L_p$ results. The $\Delta M_\text{bol}(15)$ --$t_0$ distribution is shown in Appendix~\ref{app:log_graphs}. Similar to Figure~\ref{fig:t0_LoverLp}, the observed sample (black symbols) exhibits a strong correlation between these parameters, and the models show similar qualitative behavior in both relations. As with the $ L_{30}/L_p$--$t_0$ relation, we fit a linear regression model between these parameters and find a slope of ${\approx}0.011\,\text{day}^{-1}$, about half of the value when using $L_{30}/L_p$. This result demonstrates the advantage of the $L_{30}/L_p$ parameter in distinguishing SNe. It stems from the fact that the time interval between $L_p$ and $L_{30}$ becomes smaller as the rise time increases, leading to larger values of $L_{30}/L_p$ for objects with long rise times, regardless of their decline rate. Since the observed decline rate decreases with the luminosity, the $L_{30}/L_p$ parameter changes more rapidly with respect to $t_0$ and spans a broader range of values than the $L_{p+15}/L_p$ parameter.

Explosion models are also better separated in Figure~\ref{fig:t0_LoverLp} compared to Figure~\ref{fig:t0_Lp15overLp}. In the latter, high-luminosity sub-Chandra and collision models have similar $L_{p+15}/L_p$ values despite the notable differences in their $t_0$ values. Another drawback of using $L_{p+15}$ is that the peak time of SNe Ia can slightly exceed 15 days, causing $L_{p+15}$ to correspond to times later than 30 days post-explosion when NLTE effects become significant (Figure~\ref{fig:ion_compare}). Nonetheless, calculating $L_{p+15}$ does not require explosion time, and can be easily calculated with simple interpolation, reducing the associated uncertainty compared to $L_{30}/L_p$. For completeness and to facilitate comparison with previous works, we also include $L_{p+15}/L_p$ in our analysis.


\section{The $L_{30}/L_p$--$L_p$ distribution}
\label{sec:results}

This section compares the $L_{30}/L_p$--$L_p$ distribution of observed SNe Ia to various explosion models. The results are given in Appendix~\ref{app:parameters} and shown in Figure~\ref{fig:Lp_LoverLp}. A version of Figure~\ref{fig:Lp_LoverLp} in magnitude space is provided in Appendix~\ref{app:log_graphs}.

\begin{figure*}
	\includegraphics[width=\textwidth]{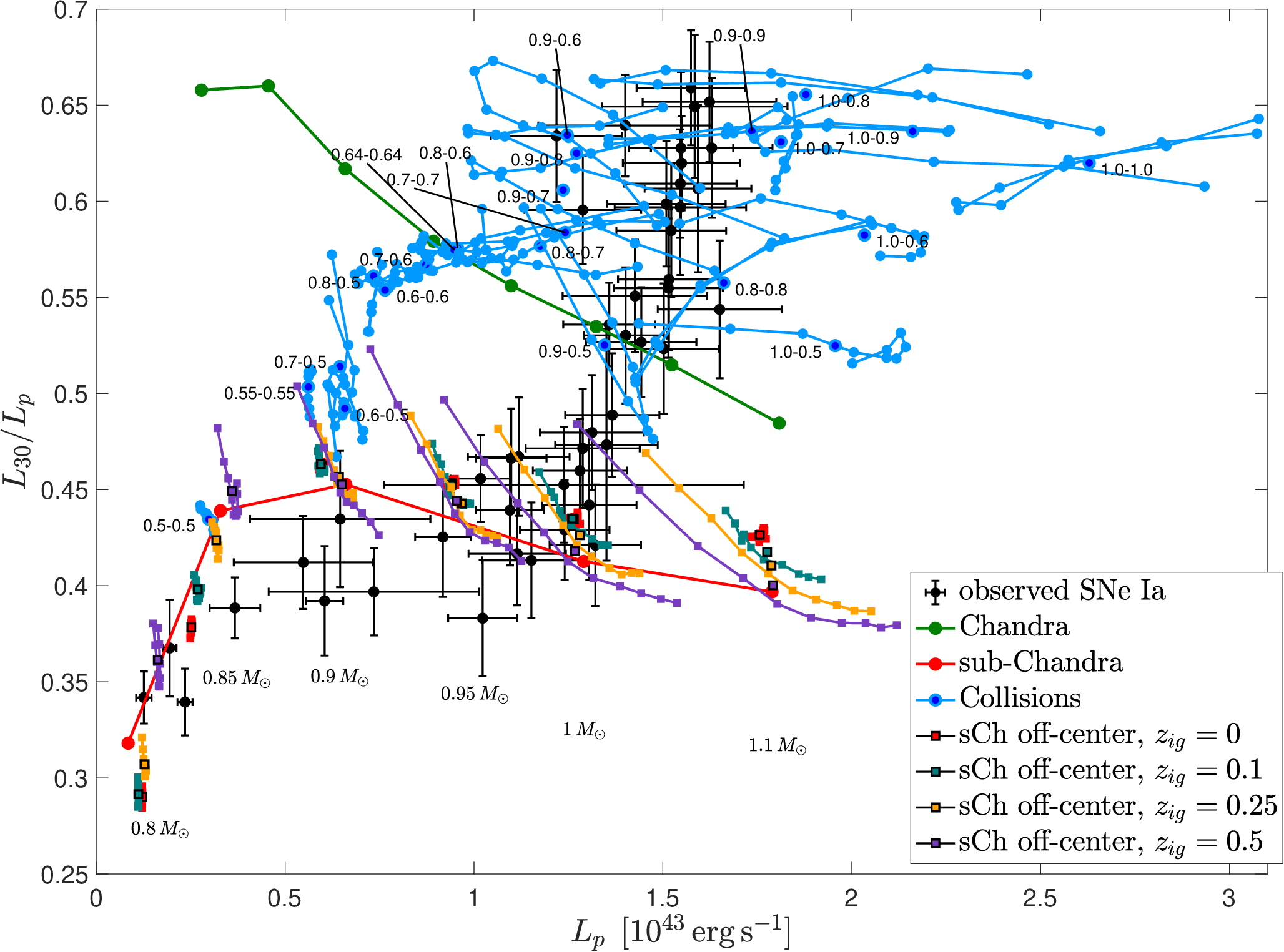}
    \caption{The $L_{30}/L_p$--$L_p$ distribution of SNe Ia, along with 1D and 2D explosion models. Observed SNe, depicted as black circles, form a tight relation, wherein $L_{30}/L_p$ increases monotonically with $L_p$ throughout the entire luminosity range. 1D models consist of the Chandra explosion of \protect\citet[][green line]{Dessart2014} and the sub-Chandra detonations of \protect\citet[][red line]{Kushnir2020}. The 2D models include direct WD collisions \protect\citep{Kushnir2013} and off-centered sub-Chandra ignitions \protect\citep{Schinasi2024}. For these models, shown are the results for different lines of sight, uniformly distributed in $\cos{\theta}$, and their angle-averaged values are indicated by a larger symbol with the same shape and a similar color, and, for the off-centered sub-Chandra models, a black-colored edge. The masses of each collision configuration are denoted in solar masses alongside their respective angle-averaged values or connected by lines. The off-centered detonations are marked according to the ignition location parameter, with red, teal, orange, and purple squares, indicating $z_\text{ig}=0,0.1,0.25,0.5$, respectively. The progenitor masses are displayed below their angle-averaged values. The $z_\text{ig}=0$ results slightly deviate from their 1D counterparts due to 2D instabilities in the thermonuclear detonation wave \citep[see][for a detailed discussion]{Schinasi2024}. None of the models can account for the observed $L_{30}/L_p$--$L_p$ distribution. A version of this figure in magnitude space is provided in Appendix~\ref{app:log_graphs}.}
    \label{fig:Lp_LoverLp}
\end{figure*}

Similar to the Phillips relation \citep{Phillips1993,Phillips1999} and to the $t_0$--$\mni$ distribution (Figure~\ref{fig:t0-Ni}), the observed bolometric width-luminosity relation shows a strong correlation between the shape and luminosity of SNe Ia, where bright SNe are slower and vice versa. Our results are broadly consistent with Figure 10 of \citet[][]{Scalzo2019}, where this trend is also noticeable. However, our sample covers a wider range of luminosities, and our relation using the $L_{30}/L_p$ parameter exhibits greater precision, within ${\approx}10$ percent in either the luminosity or $L_{30}/L_p$, compared to their relation using $\Delta M_\text{bol}(15)$, which displays a significantly more scatter, extending to tens of percents. The greater precision of our sample is partly due to the choice of the shape parameter and partly due to the data quality and analysis, as our relation using $\Delta M_\text{bol}(15)$ (see Section \ref{sec:dm15_Mp}) is tighter than the relation in \cite{Scalzo2019}.

The results of the models resemble those of the $t_0$--$\mni$ distribution (Figure~\ref{fig:t0-Ni}), where none of the models reproduce the positive correlation in the $L_{30}/L_p$--$L_p$ relation across the entire luminosity range. 

The Chandra models, represented by the green line, overlap with the observations only in a limited region. Unlike the observed relation, these models exhibit a distinct negative correlation across most luminosity ranges. The faint models are much slower than observed, while the bright models are faster.

The 1D sub-Chandra models, depicted by the red line, align well with low-luminosity SNe. However, the $L_{30}/L_p$ values for the more luminous SNe show significant tension with the observations. Notably, the more luminous SNe Ia, corresponding to \nickel\ masses of $\mni\gtrsim0.5\,M_\odot$ and $L_p\gtrsim1.2\times10^{43}\,\text{erg}\,\text{s}^{-1}$, encompass the vast majority of events \citep{Sharon2021}. The results for the off-centered ignitions of sub-Chandra WDs are shown for WD masses of $0.8,0.85,0.9,0.95,1,1.1\,M_{\odot}$ and ignition locations of $z_\text{ig}=0,0.1,0.25,0.5$. These results are presented for different values of the inclination angle $\theta$, distributed evenly in $\cos{\theta}$, along with the angle-averaged values. The $z_\text{ig}=0$ configurations slightly deviate from their 1D counterparts due to 2D instabilities in the thermonuclear detonation wave \citep[see][for a detailed discussion]{Schinasi2024}.

The models with $z_\text{ig}>0$ show significant variations due to the inclination angle, with a spread of up to ${\approx}25$ percent in $L_{30}/L_p$ and ${\approx}50$ percent in $L_p$ for the most asymmetric configurations. These variations are more pronounced for high masses, displaying a clear anti-correlation between the parameters. For $M\lesssim 0.85,M_\odot$, the observables are only marginally influenced by the inclination angle. Our results generally align with the non-spherical sub-Chandra explosions presented by \cite{Shen2021Multi}. However, their findings show larger variations in the shape parameter, particularly for low-mass configurations. Compared to the 1D sub-Chandra models, the off-centered detonation results cover a larger fraction of the $L_{30}/L_p$--$L_p$ distribution and account for some of the observed spread. Nonetheless, they fail to reproduce the $L_{30}/L_p$--$L_p$ positive correlation for the luminous SNe, where the predicted $L_{30}/L_p-L_p$ is smaller by ${\approx}25$ percent compared to observations.


The head-on WD collisions, observed at various inclination angles evenly distributed in $\cos{\theta}$, are represented by blue lines and circles, with dark blue circles indicating the angle-averaged values. The masses of each configuration are denoted in solar masses alongside their respective angle-averaged values or connected by lines. Configurations with equal masses exhibit reflection symmetry with respect to the plane perpendicular to the symmetry axis. Consequently, observables at angles $\theta$ and $-\theta$ are expected to align. Figure~\ref{fig:Lp_LoverLp} confirms this alignment, albeit with slight variations due to Monte Carlo noise. These models can reproduce only the bright segment of the observed distribution. While models with intermediate masses achieve luminosities and shape parameters roughly consistent with observations, their luminosity varies significantly across different $\theta$ values, often exceeding observed luminosities. The shape parameter of these models is less sensitive to variations in the observed angle. The low-luminosity models exhibit high values of $L_{30}/L_p$ compared to observations at similar luminosities and fail to match the observed relation across all $\theta$ values. In these models, the observed inclination angle primarily affects $L_{30}/L_p$, altering it by up to ${\approx}20$ percent, which is insufficient to agree with the observed values. The relationship between the observed angle variations and the ejecta properties is a topic for future study.


In summary, none of the models can account for the observed $L_{30}/L_p$--$L_p$ distribution.


\subsection{The $L_{p+15}/L_p$--$L_p$ relation}
\label{sec:dm15_Mp}



\begin{figure}
     \centering
     \includegraphics[width=\columnwidth]{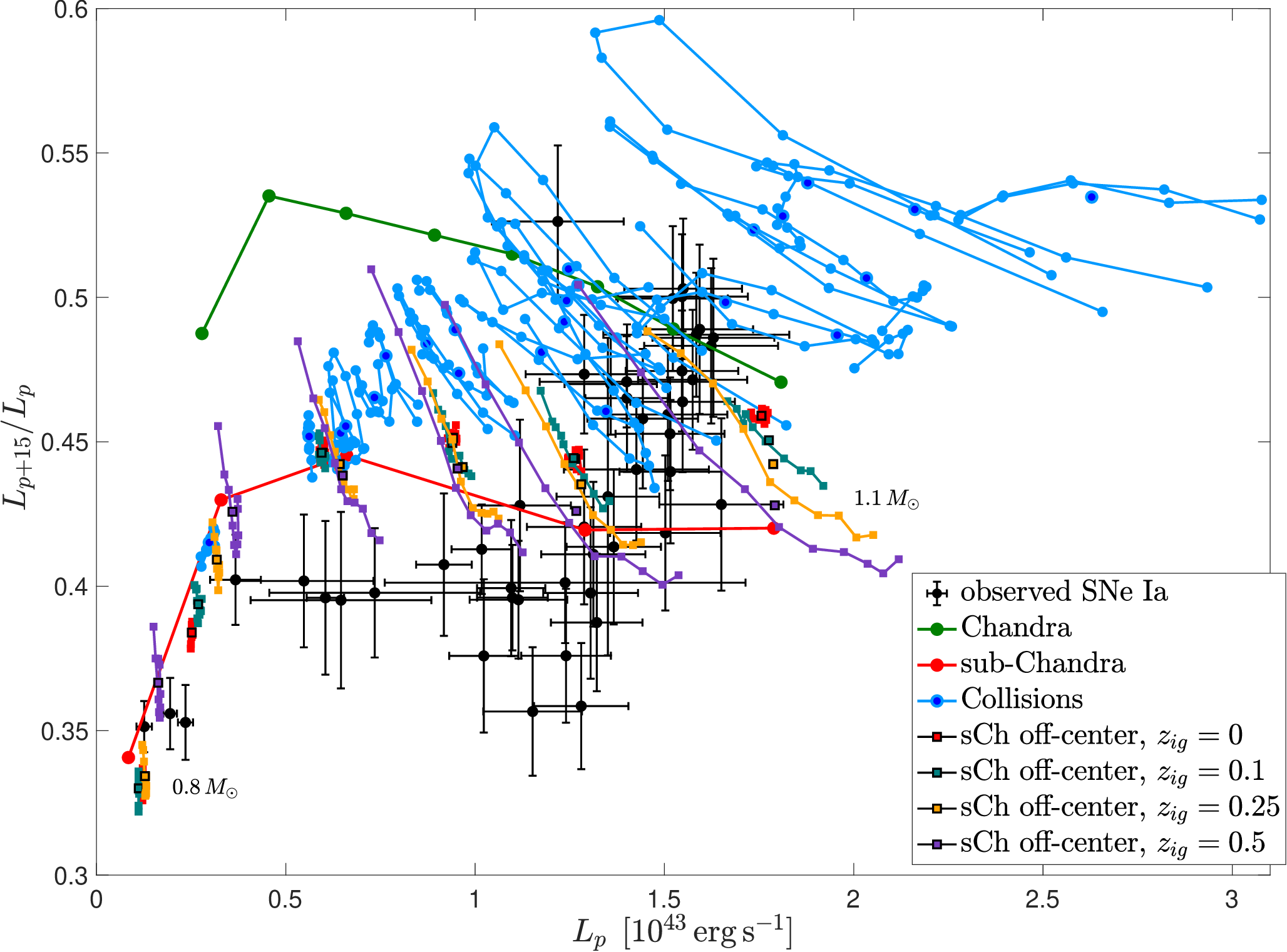}
     \caption{Same as Figure~\ref{fig:Lp_LoverLp}, but for the $L_{p+15}/L_p$--$L_p$ distribution. Compared to Figure~\ref{fig:Lp_LoverLp}, the $L_{p+15}/L_p$ shape parameter covers a smaller range of values and is not strictly monotonic with the luminosity. Additionally, the models are not as well separated, which is most evident in the overlap between the sub-Chandrasekhar and collision models. A version of this figure in magnitude space is provided in Appendix~\ref{app:log_graphs}.}
     \label{fig:Lp_Lp15overLp}
 \end{figure}

In Figure~\ref{fig:Lp_Lp15overLp}, we present the observed distribution of $L_{p+15}/L_p$ and $L_p$. The general trend that lower luminosity SNe Ia evolve more rapidly is evident in this relationship. However, as noted in Section~\ref{sec:Selecting}, the  $L_{p+15}/L_p$ parameter covers a smaller range of values, and the relation is not strictly monotonic, with some intermediate-luminosity objects displaying very high shape parameters.

The models' behavior mirrors that of the $L_{30}/L_p$--$L_p$ relation, but the separation between the models is less pronounced. This is most apparent in the overlap between the high-luminosity sub-Chandrasekhar models and the collision models, which is also evident in the $L_{p+15}/L_p$--$t_0$ relation (Figure~\ref{fig:t0_Lp15overLp}). Here, sub-Chandrasekhar models exhibit $L_{p+15}/L_p$ values that are higher than observed for SNe with low to intermediate luminosities, whereas the $L_{30}/L_p$--$L_p$ relation better represented these SNe with the sub-Chandrasekhar models. Conversely, sub-Chandrasekhar models align more closely with the high luminosity regime, where most SNe Ia are situated. Results for viewing angles opposite the explosion point of more massive progenitors align with the most luminous objects. However, for viewing angles near the explosion point, the models predict too high luminosities that evolve more rapidly than observed data. A version of this figure in magnitude space is provided in Appendix~\ref{app:log_graphs}.

For the Chandra models, our findings are broadly consistent with those of \cite{Kasen2007}, who derived light curves for Chandrasekhar-mass WD explosions and observed nearly constant $\Delta M_\text{bol}(15)$ values across a $\mni$ range of $0.35$--$0.7\,M_\odot$. In this luminosity range, our $\Delta M_\text{bol}(15)$ results exhibit similar values but show variations of ${\approx}15$ percent.

\section{Opacity of low-luminosity models}
\label{sec:opacity}

We demonstrated in Section~\ref{sec:shape_parameter} that the sub-Chandra models exhibit a peculiar behavior in the low-luminosity region regarding the two shape parameters considered in this work, $t_0$ and $L_{30}/L_p$.  While $t_0$ increases monotonically as the luminosity decreases, $L_{30}/L_p$ does not follow a monotonic trend across all luminosities and reverses for the two faintest models. This anti-correlation is evident in Figure~\ref{fig:t0_LoverLp}, where $L_{30}/L_p$ of the sub-Chandra models drops for high values of $t_0$, corresponding to the fainter models of the $0.8$ and $0.85\,M_\odot$ WD progenitors. The drop in $L_{30}/L_p$ is attributed to the short rise time evident in these models.
Since both the rise time and the $\gamma$-ray escape time are expected to increase with the ejecta's mass and decrease with the kinetic energy of the ejecta and the extent of the \nickel\ distribution within it, this suggests that the anti-correlation is likely due to opacity differences. 

As discussed in the preceding sections, the opacity of optical light, which depends on the temperature, density, and composition of the matter, significantly impacts the evolution of the SNe light curve. In contrast, the opacity for $\gamma$-ray transfer is almost constant for the ejecta considered here. Therefore, we evaluate and compare the characteristic optical opacity of the models. This is not straightforward, as optical opacity is wavelength-dependent, varies throughout the ejecta, and evolves over time. To address the wavelength dependency, we use the Rosseland mean opacity since the photons travel through diffusion during the early stages of the SN. For each epoch, we calculate the mass-averaged Rosseland mean opacity:
\begin{equation}
\label{eq:mean_opacity}
    \left<\kappa_\textrm{R}(t)\right> = \frac{1}{m_*}\int_0^{m_*(t)}\kappa_\text{R}dm,
\end{equation}
where $m_*(t)$ is the enclosed mass within which the diffusion approximation is valid, determined by an optical depth from the ejecta outer edge of $\tau>2$.

\begin{figure}
	\includegraphics[width=\columnwidth]{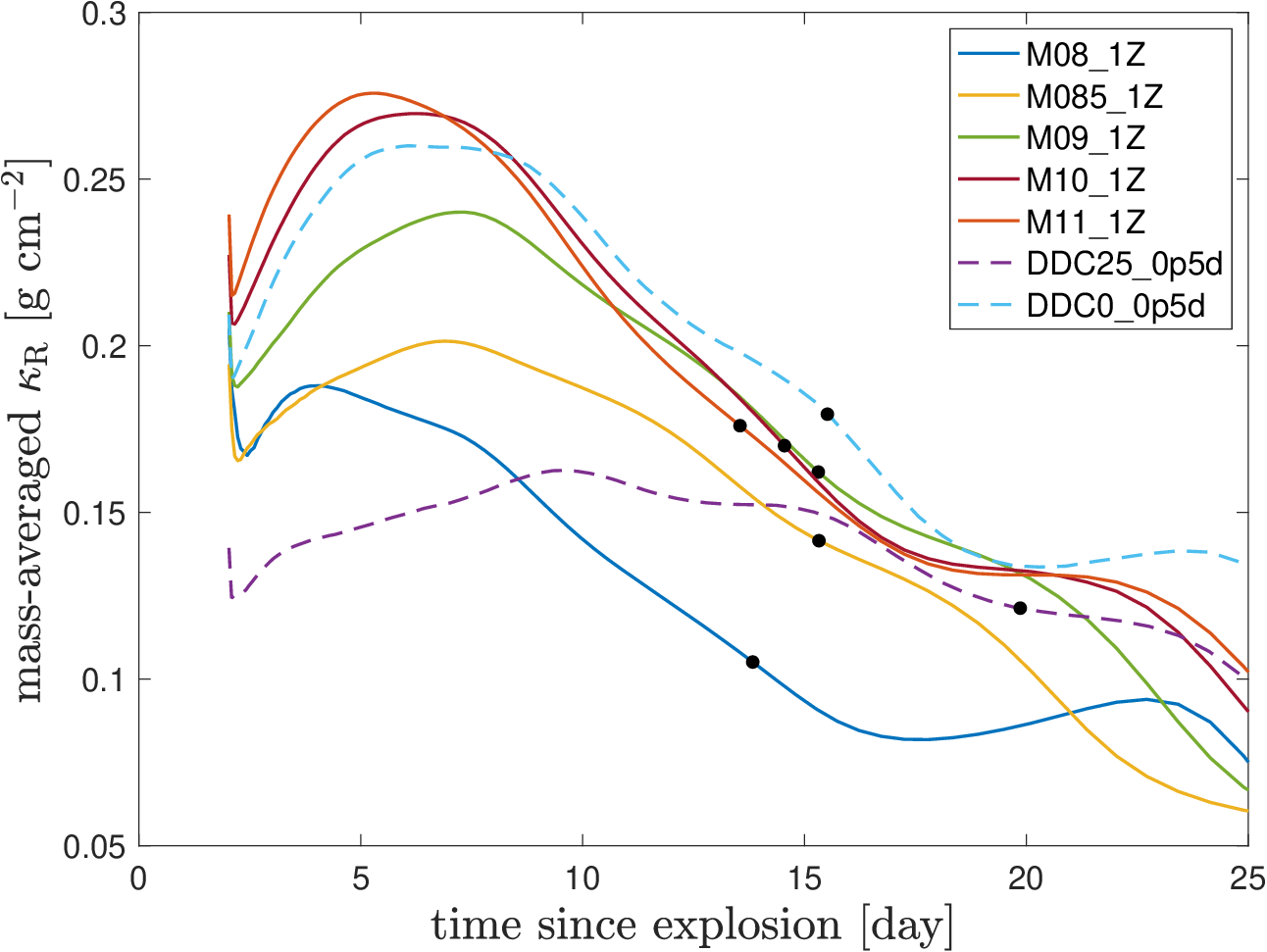}
    \caption{Mass-averaged Rosseland mean opacity as a function of time for several selected models. The models include the 1D sub-Chandra configurations (solid lines) and the two Chandra models with the faintest and brightest luminosities (dashed lines). The black point on each curve indicates the epoch of peak light. Averaging was performed over all masses with an optical depth from the outer edge of the ejecta satisfying $\tau>2$. The opacity of the $0.8\,M_\odot$ WD sub-Chandra model is significantly lower than the other models, consistent with its short rise time.}
    \label{fig:kappa_t}
\end{figure}

The temporal evolution of the mass-averaged Rosseland mean opacity for several selected models is shown in Figure~\ref{fig:kappa_t}, with the epoch of peak light for each model marked by a black point. The sub-Chandra models, represented by solid lines, exhibit a significant decrease in average opacity for the low-mass progenitors compared to other models. This decrease is particularly pronounced in the $0.8\,M_\odot$ model, which has substantially lower opacities -- nearly half as much around peak light -- compared to the other models. The opacities of the faintest and brightest Chandra models, indicated by the dashed lines, also show significant variation. The low opacity of the faint models can be attributed to the differences in the composition of these models, namely the amount of iron-group elements, or to the conditions within the ejecta during the evolution of the SN, such as the temperature. We repeated this calculation for the $0.8\,M_\odot$ sub-Chandra model, altering the composition to include additional stable iron-group isotopes, and found that the Rosseland mean opacity does not change significantly. We therefore attribute the low values of opacity to the temperature in the ejecta, which is considerably lower for these low-luminosity models, and has a substantial effect on the opacity. However, a more thorough analysis is beyond the scope of this paper.


The opacity variations observed for both the sub-Chandra and Chandra models are consistent with their behavior in Figure~\ref{fig:t0_LoverLp}. These variations explain the short rise times of the low-luminosity sub-Chandra models despite their relatively high $t_0$ values. In Appendix~\ref{app:opacity}, we provide a quantitative analysis of these results and directly relate the models' rise time to their average Rosseland mean opacity.

\section{Summary and Conclusions}
\label{sec:discussion}

In this study, we aimed to identify an observational width--luminosity relation, similar to the $t_0$--$M_\mathrm{Ni56}$ relation, that would allow to constrain multi-D models at pre-nebular phases while minimizing the inherent uncertainties of RT calculations. We argued that the bolometric light curve up to 30 days from the explosion can be computed without accounting for NLTE effects, which are inherently uncertain and difficult to compute (Section~\ref{sec:shape_parameter}). Subsequently, we introduced a width-luminosity relation for SNe Ia, where the peak luminosity $L_p$ is compared with the ratio of the luminosity at 30 days from the explosion to the peak luminosity, $L_{30}/L_p$. The distribution of these parameters for a sample of 47 SNe Ia is presented in Figure~\ref{fig:Lp_LoverLp}, along with the results from 1D and 2D explosion models (see Section~\ref{sec:results}).

In previous work, we compared the observed $t_0$--$M_\mathrm{Ni56}$ relation to known SNe Ia models from the literature \citep[][also see Figure~\ref{fig:t0-Ni} for an updated relation]{Sharon2020b}. Although that analysis bypassed RT calculations, it could not account for multi-D effects and required stringent observational conditions, including late-time observations up to ${\sim}100$ days. While the current analysis relies on RT calculations, it does not face these limitations. The conclusion of this work aligns with that of \cite{Sharon2020b}: None of the known SNe Ia models can reproduce the observed width-luminosity distribution of SNe Ia, even when considering multi-D effects.

Several avenues could potentially align the models more closely with the observations. One possibility involves adjusting the initial composition of the sub-Chandra WDs, which evolve too rapidly compared to the bulk of the observations, from CO towards heavier elements. This adjustment would reduce the available thermonuclear energy of the WD, resulting in lower velocities of the ejected material and a slower evolution of the light curve. However, a heavier composition is expected only for very massive WDs \citep[$M\gtrsim1.1,M_\odot$]{Lauffer2018}. We are currently investigating this possibility, and the findings will be published in a subsequent work.

Another avenue is to perform more accurate calculations of the collision model. The calculations by \cite{Kushnir2013} were carried out with relatively low resolution and a simplified 13-isotope reaction network. Preliminary results from more accurate calculations do not show a substantial change in the $L_{30}/L_p\text{--}L_p$ distribution and will be reported in a future paper. Additionally, 3D calculations of collisions with non-zero impact parameters could yield different results; however, such high-resolution calculations are currently beyond our computational capabilities.

Throughout the paper, we have mentioned non-thermal processes as significant sources of uncertainty in RT modeling. These processes substantially impact the SN's evolution, yet their full extent and impact remain poorly understood. Traditionally, non-thermal effects are primarily considered during the nebular phase; however, \cite{Dessart2012} demonstrated that non-thermal excitation alters the spectra of Type Ib SNe, even at earlier phases, during the photospheric phase. Therefore, it is crucial to understand their impact on pre-nebular times and the SN light curves. Furthermore, parameters associated with non-thermal processes contain considerable uncertainties. For example, electron-impact ionization cross-sections are often unknown, and the absence of such cross-sections for Co III ionization in CMFGEN causes substantial discrepancies in Co and Fe ionization levels, as illustrated in Figure~\ref{fig:ion_compare}. Future work should study in more detail the effect non-thermal processes have on the SNe observables and their sensitivity to the non-thermal parameters.

\section*{Acknowledgements}

We thank Boaz Katz for offering this work and for the useful discussions. We thank Stéphane Blondin and John Hillier for their assistance regarding CMFGEN. DK is supported by a research grant from The Abramson Family Center for Young Scientists, and by the Minerva Stiftung.

\section*{Data Availability}

The data underlying this article are available in the article and in its online supplementary material.



\bibliographystyle{mnras}
\bibliography{bibliography} 

\newcommand{\noop}[1]{}
\begin{thebibliography}{}
\makeatletter
\relax
\def\mn@urlcharsother{\let\do\@makeother \do\$\do\&\do\#\do\^\do\_\do\%\do\~}
\def\mn@doi{\begingroup\mn@urlcharsother \@ifnextchar [ {\mn@doi@} {\mn@doi@[]}}
\def\mn@doi@[#1]#2{\def\@tempa{#1}\ifx\@tempa\@empty \href {http://dx.doi.org/#2} {doi:#2}\else \href {http://dx.doi.org/#2} {#1}\fi \endgroup}
\def\mn@eprint#1#2{\mn@eprint@#1:#2::\@nil}
\def\mn@eprint@arXiv#1{\href {http://arxiv.org/abs/#1} {{\tt arXiv:#1}}}
\def\mn@eprint@dblp#1{\href {http://dblp.uni-trier.de/rec/bibtex/#1.xml} {dblp:#1}}
\def\mn@eprint@#1:#2:#3:#4\@nil{\def\@tempa {#1}\def\@tempb {#2}\def\@tempc {#3}\ifx \@tempc \@empty \let \@tempc \@tempb \let \@tempb \@tempa \fi \ifx \@tempb \@empty \def\@tempb {arXiv}\fi \@ifundefined {mn@eprint@\@tempb}{\@tempb:\@tempc}{\expandafter \expandafter \csname mn@eprint@\@tempb\endcsname \expandafter{\@tempc}}}

\bibitem[\protect\citeauthoryear{{Arnaud} \& {Rothenflug}}{{Arnaud} \& {Rothenflug}}{1985}]{Arnaud1985}
{Arnaud} M.,  {Rothenflug} R.,  1985, \aaps, \href {https://ui.adsabs.harvard.edu/abs/1985A&AS...60..425A} {60, 425}

\bibitem[\protect\citeauthoryear{{Axelrod}}{{Axelrod}}{1980}]{Axelrod1980}
{Axelrod} T.~S.,  1980, PhD thesis, University of California, Santa Cruz

\bibitem[\protect\citeauthoryear{{Baron}, {Hauschildt}, {Nugent}  \& {Branch}}{{Baron} et~al.}{1996}]{Baron1996}
{Baron} E.,  {Hauschildt} P.~H.,  {Nugent} P.,   {Branch} D.,  1996, \mn@doi [\mnras] {10.1093/mnras/283.1.297}, \href {https://ui.adsabs.harvard.edu/abs/1996MNRAS.283..297B} {283, 297}

\bibitem[\protect\citeauthoryear{{Blinnikov} \& {Bartunov}}{{Blinnikov} \& {Bartunov}}{1993}]{Blinnikov1993}
{Blinnikov} S.~I.,  {Bartunov} O.~S.,  1993, \aap, \href {https://ui.adsabs.harvard.edu/abs/1993A&A...273..106B} {273, 106}

\bibitem[\protect\citeauthoryear{{Blondin}, {Dessart}, {Hillier}  \& {Khokhlov}}{{Blondin} et~al.}{2017}]{Blondin2017}
{Blondin} S.,  {Dessart} L.,  {Hillier} D.~J.,   {Khokhlov} A.~M.,  2017, \mn@doi [\mnras] {10.1093/mnras/stw2492}, \href {https://ui.adsabs.harvard.edu/abs/2017MNRAS.470..157B} {470, 157}

\bibitem[\protect\citeauthoryear{{Blondin} et~al.,}{{Blondin} et~al.}{2022}]{Blondin2022}
{Blondin} S.,  et~al., 2022, \mn@doi [\aap] {10.1051/0004-6361/202244134}, \href {https://ui.adsabs.harvard.edu/abs/2022A&A...668A.163B} {668, A163}

\bibitem[\protect\citeauthoryear{{Burns} et~al.,}{{Burns} et~al.}{2018}]{Burns2018}
{Burns} C.~R.,  et~al., 2018, \mn@doi [\apj] {10.3847/1538-4357/aae51c}, \href {https://ui.adsabs.harvard.edu/abs/2018ApJ...869...56B} {869, 56}

\bibitem[\protect\citeauthoryear{{Chen} et~al.,}{{Chen} et~al.}{2023}]{Chen2023}
{Chen} N.~M.,  et~al., 2023, \mn@doi [\apjl] {10.3847/2041-8213/acb6d8}, \href {https://ui.adsabs.harvard.edu/abs/2023ApJ...944L..28C} {944, L28}

\bibitem[\protect\citeauthoryear{{Collins}, {Gronow}, {Sim}  \& {R{\"o}pke}}{{Collins} et~al.}{2022}]{Collins2022}
{Collins} C.~E.,  {Gronow} S.,  {Sim} S.~A.,   {R{\"o}pke} F.~K.,  2022, \mn@doi [\mnras] {10.1093/mnras/stac2665}, \href {https://ui.adsabs.harvard.edu/abs/2022MNRAS.517.5289C} {517, 5289}

\bibitem[\protect\citeauthoryear{{Contardo}, {Leibundgut}  \& {Vacca}}{{Contardo} et~al.}{2000}]{Contardo2000}
{Contardo} G.,  {Leibundgut} B.,   {Vacca} W.~D.,  2000, \mn@doi [\aap] {10.48550/arXiv.astro-ph/0005507}, \href {https://ui.adsabs.harvard.edu/abs/2000A&A...359..876C} {359, 876}

\bibitem[\protect\citeauthoryear{{Contreras} et~al.,}{{Contreras} et~al.}{2010}]{Contreras2010}
{Contreras} C.,  et~al., 2010, \mn@doi [\aj] {10.1088/0004-6256/139/2/519}, \href {https://ui.adsabs.harvard.edu/abs/2010AJ....139..519C} {139, 519}

\bibitem[\protect\citeauthoryear{{Dessart} \& {Hillier}}{{Dessart} \& {Hillier}}{2010}]{Dessart2010}
{Dessart} L.,  {Hillier} D.~J.,  2010, \mn@doi [\mnras] {10.1111/j.1365-2966.2010.16611.x}, \href {https://ui.adsabs.harvard.edu/abs/2010MNRAS.405.2141D} {405, 2141}

\bibitem[\protect\citeauthoryear{{Dessart}, {Hillier}, {Li}  \& {Woosley}}{{Dessart} et~al.}{2012}]{Dessart2012}
{Dessart} L.,  {Hillier} D.~J.,  {Li} C.,   {Woosley} S.,  2012, \mn@doi [\mnras] {10.1111/j.1365-2966.2012.21374.x}, \href {https://ui.adsabs.harvard.edu/abs/2012MNRAS.424.2139D} {424, 2139}

\bibitem[\protect\citeauthoryear{{Dessart}, {Blondin}, {Hillier}  \& {Khokhlov}}{{Dessart} et~al.}{2014a}]{Dessart2014}
{Dessart} L.,  {Blondin} S.,  {Hillier} D.~J.,   {Khokhlov} A.,  2014a, \mn@doi [\mnras] {10.1093/mnras/stu598}, \href {https://ui.adsabs.harvard.edu/abs/2014MNRAS.441..532D} {441, 532}

\bibitem[\protect\citeauthoryear{{Dessart}, {Hillier}, {Blondin}  \& {Khokhlov}}{{Dessart} et~al.}{2014b}]{Dessart2014radiative}
{Dessart} L.,  {Hillier} D.~J.,  {Blondin} S.,   {Khokhlov} A.,  2014b, \mn@doi [\mnras] {10.1093/mnras/stu789}, \href {https://ui.adsabs.harvard.edu/abs/2014MNRAS.441.3249D} {441, 3249}

\bibitem[\protect\citeauthoryear{{Ganeshalingam}, {Li}  \& {Filippenko}}{{Ganeshalingam} et~al.}{2011}]{Ganeshalingam2011}
{Ganeshalingam} M.,  {Li} W.,   {Filippenko} A.~V.,  2011, \mn@doi [\mnras] {10.1111/j.1365-2966.2011.19213.x}, \href {https://ui.adsabs.harvard.edu/abs/2011MNRAS.416.2607G} {416, 2607}

\bibitem[\protect\citeauthoryear{{Gronow}, {Collins}, {Sim}  \& {R{\"o}pke}}{{Gronow} et~al.}{2021}]{Gronow2021}
{Gronow} S.,  {Collins} C.~E.,  {Sim} S.~A.,   {R{\"o}pke} F.~K.,  2021, \mn@doi [\aap] {10.1051/0004-6361/202039954}, \href {https://ui.adsabs.harvard.edu/abs/2021A&A...649A.155G} {649, A155}

\bibitem[\protect\citeauthoryear{{Guttman}, {Shenhar}, {Sarkar}  \& {Waxman}}{{Guttman} et~al.}{2024}]{Guttman2024}
{Guttman} O.,  {Shenhar} B.,  {Sarkar} A.,   {Waxman} E.,  2024, \mn@doi [arXiv e-prints] {10.48550/arXiv.2403.08769}, \href {https://ui.adsabs.harvard.edu/abs/2024arXiv240308769G} {p. arXiv:2403.08769}

\bibitem[\protect\citeauthoryear{{Hillebrandt} \& {Niemeyer}}{{Hillebrandt} \& {Niemeyer}}{2000}]{Hillebrandt2000}
{Hillebrandt} W.,  {Niemeyer} J.~C.,  2000, \mn@doi [\araa] {10.1146/annurev.astro.38.1.191}, \href {https://ui.adsabs.harvard.edu/abs/2000ARA%26A..38..191H} {38, 191}

\bibitem[\protect\citeauthoryear{{Hillier} \& {Miller}}{{Hillier} \& {Miller}}{1998}]{Hillier1998}
{Hillier} D.~J.,  {Miller} D.~L.,  1998, \mn@doi [\apj] {10.1086/305350}, \href {https://ui.adsabs.harvard.edu/abs/1998ApJ...496..407H} {496, 407}

\bibitem[\protect\citeauthoryear{Jeffery}{Jeffery}{1999}]{Jeffery1999}
Jeffery D.~J.,  1999, arXiv preprint astro-ph/9907015

\bibitem[\protect\citeauthoryear{{Jerkstrand}}{{Jerkstrand}}{2011}]{Jerkstrand2011PhD}
{Jerkstrand} A.,  2011, PhD thesis, Stockholm University

\bibitem[\protect\citeauthoryear{{Kasen}}{{Kasen}}{2006}]{Kasen2006}
{Kasen} D.,  2006, \mn@doi [\apj] {10.1086/506588}, \href {https://ui.adsabs.harvard.edu/abs/2006ApJ...649..939K} {649, 939}

\bibitem[\protect\citeauthoryear{{Kasen} \& {Woosley}}{{Kasen} \& {Woosley}}{2007}]{Kasen2007}
{Kasen} D.,  {Woosley} S.~E.,  2007, \mn@doi [\apj] {10.1086/510375}, \href {https://ui.adsabs.harvard.edu/abs/2007ApJ...656..661K} {656, 661}

\bibitem[\protect\citeauthoryear{{Kasen}, {Thomas}  \& {Nugent}}{{Kasen} et~al.}{2006}]{Kasen2006Sedona}
{Kasen} D.,  {Thomas} R.~C.,   {Nugent} P.,  2006, \mn@doi [\apj] {10.1086/506190}, \href {https://ui.adsabs.harvard.edu/abs/2006ApJ...651..366K} {651, 366}

\bibitem[\protect\citeauthoryear{{Kozma} \& {Fransson}}{{Kozma} \& {Fransson}}{1992}]{Kozma1992}
{Kozma} C.,  {Fransson} C.,  1992, \mn@doi [\apj] {10.1086/171311}, \href {https://ui.adsabs.harvard.edu/abs/1992ApJ...390..602K} {390, 602}

\bibitem[\protect\citeauthoryear{{Krisciunas} et~al.,}{{Krisciunas} et~al.}{2017}]{Krisciunas2017}
{Krisciunas} K.,  et~al., 2017, \mn@doi [\aj] {10.3847/1538-3881/aa8df0}, \href {https://ui.adsabs.harvard.edu/abs/2017AJ....154..211K} {154, 211}

\bibitem[\protect\citeauthoryear{{Kushnir} \& {Katz}}{{Kushnir} \& {Katz}}{2019}]{KushnirKatz2019}
{Kushnir} D.,  {Katz} B.,  2019, \mn@doi [Research Notes of the American Astronomical Society] {10.3847/2515-5172/ab5064}, \href {https://ui.adsabs.harvard.edu/abs/2019RNAAS...3..162K} {3, 162}

\bibitem[\protect\citeauthoryear{{Kushnir}, {Katz}, {Dong}, {Livne}  \& {Fern{\'a}ndez}}{{Kushnir} et~al.}{2013}]{Kushnir2013}
{Kushnir} D.,  {Katz} B.,  {Dong} S.,  {Livne} E.,   {Fern{\'a}ndez} R.,  2013, \mn@doi [\apjl] {10.1088/2041-8205/778/2/L37}, \href {https://ui.adsabs.harvard.edu/abs/2013ApJ...778L..37K} {778, L37}

\bibitem[\protect\citeauthoryear{{Kushnir}, {Wygoda}  \& {Sharon}}{{Kushnir} et~al.}{2020}]{Kushnir2020}
{Kushnir} D.,  {Wygoda} N.,   {Sharon} A.,  2020, \mn@doi [\mnras] {10.1093/mnras/staa3017}, \href {https://ui.adsabs.harvard.edu/abs/2020MNRAS.499.4725K} {499, 4725}

\bibitem[\protect\citeauthoryear{{Kwok} et~al.,}{{Kwok} et~al.}{2023}]{Kwok2023}
{Kwok} L.~A.,  et~al., 2023, \mn@doi [\apjl] {10.3847/2041-8213/acb4ec}, \href {https://ui.adsabs.harvard.edu/abs/2023ApJ...944L...3K} {944, L3}

\bibitem[\protect\citeauthoryear{{Lauffer}, {Romero}  \& {Kepler}}{{Lauffer} et~al.}{2018}]{Lauffer2018}
{Lauffer} G.~R.,  {Romero} A.~D.,   {Kepler} S.~O.,  2018, \mn@doi [\mnras] {10.1093/mnras/sty1925}, \href {https://ui.adsabs.harvard.edu/abs/2018MNRAS.480.1547L} {480, 1547}

\bibitem[\protect\citeauthoryear{{Li}, {Hillier}  \& {Dessart}}{{Li} et~al.}{2012}]{Li2012}
{Li} C.,  {Hillier} D.~J.,   {Dessart} L.,  2012, \mn@doi [\mnras] {10.1111/j.1365-2966.2012.21198.x}, \href {https://ui.adsabs.harvard.edu/abs/2012MNRAS.426.1671L} {426, 1671}

\bibitem[\protect\citeauthoryear{{Maoz}, {Mannucci}  \& {Nelemans}}{{Maoz} et~al.}{2014}]{Maoz2014}
{Maoz} D.,  {Mannucci} F.,   {Nelemans} G.,  2014, \mn@doi [\araa] {10.1146/annurev-astro-082812-141031}, \href {https://ui.adsabs.harvard.edu/abs/2014ARA&A..52..107M} {52, 107}

\bibitem[\protect\citeauthoryear{{Nahar}}{{Nahar}}{1996}]{Nahar1996}
{Nahar} S.~N.,  1996, \mn@doi [\pra] {10.1103/PhysRevA.53.2417}, \href {https://ui.adsabs.harvard.edu/abs/1996PhRvA..53.2417N} {53, 2417}

\bibitem[\protect\citeauthoryear{{Noebauer} \& {Sim}}{{Noebauer} \& {Sim}}{2019}]{Noebauer2019}
{Noebauer} U.~M.,  {Sim} S.~A.,  2019, \mn@doi [Living Reviews in Computational Astrophysics] {10.1007/s41115-019-0004-9}, \href {https://ui.adsabs.harvard.edu/abs/2019LRCA....5....1N} {5, 1}

\bibitem[\protect\citeauthoryear{{Phillips}}{{Phillips}}{1993}]{Phillips1993}
{Phillips} M.~M.,  1993, \mn@doi [\apjl] {10.1086/186970}, \href {https://ui.adsabs.harvard.edu/abs/1993ApJ...413L.105P} {413, L105}

\bibitem[\protect\citeauthoryear{{Phillips}, {Lira}, {Suntzeff}, {Schommer}, {Hamuy}  \& {Maza}}{{Phillips} et~al.}{1999}]{Phillips1999}
{Phillips} M.~M.,  {Lira} P.,  {Suntzeff} N.~B.,  {Schommer} R.~A.,  {Hamuy} M.,   {Maza} J.,  1999, \mn@doi [\aj] {10.1086/301032}, \href {https://ui.adsabs.harvard.edu/abs/1999AJ....118.1766P} {118, 1766}

\bibitem[\protect\citeauthoryear{{Phillips} et~al.,}{{Phillips} et~al.}{2006}]{Phillips2006}
{Phillips} M.~M.,  et~al., 2006, \mn@doi [\aj] {10.1086/503108}, \href {https://ui.adsabs.harvard.edu/abs/2006AJ....131.2615P} {131, 2615}

\bibitem[\protect\citeauthoryear{{Pinto} \& {Eastman}}{{Pinto} \& {Eastman}}{2000}]{PintoEastman2000}
{Pinto} P.~A.,  {Eastman} R.~G.,  2000, \mn@doi [\apj] {10.1086/308380}, \href {https://ui.adsabs.harvard.edu/abs/2000ApJ...530..757P} {530, 757}

\bibitem[\protect\citeauthoryear{{Riess} et~al.,}{{Riess} et~al.}{2022}]{Riess2022}
{Riess} A.~G.,  et~al., 2022, \mn@doi [\apjl] {10.3847/2041-8213/ac5c5b}, \href {https://ui.adsabs.harvard.edu/abs/2022ApJ...934L...7R} {934, L7}

\bibitem[\protect\citeauthoryear{{Scalzo} et~al.,}{{Scalzo} et~al.}{2014}]{Scalzo2014}
{Scalzo} R.,  et~al., 2014, \mn@doi [\mnras] {10.1093/mnras/stu350}, \href {https://ui.adsabs.harvard.edu/abs/2014MNRAS.440.1498S} {440, 1498}

\bibitem[\protect\citeauthoryear{{Scalzo} et~al.,}{{Scalzo} et~al.}{2019}]{Scalzo2019}
{Scalzo} R.~A.,  et~al., 2019, \mn@doi [\mnras] {10.1093/mnras/sty3178}, \href {https://ui.adsabs.harvard.edu/abs/2019MNRAS.483..628S} {483, 628}

\bibitem[\protect\citeauthoryear{{Schinasi-Lemberg} \& {Kushnir}}{{Schinasi-Lemberg} \& {Kushnir}}{2024}]{Schinasi2024}
{Schinasi-Lemberg} E.,  {Kushnir} D.,  2024, in preparation

\bibitem[\protect\citeauthoryear{{Sharon} \& {Kushnir}}{{Sharon} \& {Kushnir}}{2020a}]{Sharon2020b}
{Sharon} A.,  {Kushnir} D.,  2020a, \mn@doi [Research Notes of the American Astronomical Society] {10.3847/2515-5172/abb9a3}, \href {https://ui.adsabs.harvard.edu/abs/2020RNAAS...4..158S} {4, 158}

\bibitem[\protect\citeauthoryear{{Sharon} \& {Kushnir}}{{Sharon} \& {Kushnir}}{2020b}]{Sharon2020}
{Sharon} A.,  {Kushnir} D.,  2020b, \mn@doi [\mnras] {10.1093/mnras/staa1745}, \href {https://ui.adsabs.harvard.edu/abs/2020MNRAS.496.4517S} {496, 4517}

\bibitem[\protect\citeauthoryear{{Sharon} \& {Kushnir}}{{Sharon} \& {Kushnir}}{2022}]{Sharon2021}
{Sharon} A.,  {Kushnir} D.,  2022, \mn@doi [\mnras] {10.1093/mnras/stab3380}, \href {https://ui.adsabs.harvard.edu/abs/2022MNRAS.509.5275S} {509, 5275}

\bibitem[\protect\citeauthoryear{{Sharon} \& {Kushnir}}{{Sharon} \& {Kushnir}}{2023}]{Sharon2023}
{Sharon} A.,  {Kushnir} D.,  2023, \mn@doi [\mnras] {10.1093/mnras/stad1227}, \href {https://ui.adsabs.harvard.edu/abs/2023MNRAS.522.6264S} {522, 6264}

\bibitem[\protect\citeauthoryear{{Shen}, {Blondin}, {Kasen}, {Dessart}, {Townsley}, {Boos}  \& {Hillier}}{{Shen} et~al.}{2021a}]{Shen2021NLTE}
{Shen} K.~J.,  {Blondin} S.,  {Kasen} D.,  {Dessart} L.,  {Townsley} D.~M.,  {Boos} S.,   {Hillier} D.~J.,  2021a, \mn@doi [\apjl] {10.3847/2041-8213/abe69b}, \href {https://ui.adsabs.harvard.edu/abs/2021ApJ...909L..18S} {909, L18}

\bibitem[\protect\citeauthoryear{{Shen}, {Boos}, {Townsley}  \& {Kasen}}{{Shen} et~al.}{2021b}]{Shen2021Multi}
{Shen} K.~J.,  {Boos} S.~J.,  {Townsley} D.~M.,   {Kasen} D.,  2021b, \mn@doi [\apj] {10.3847/1538-4357/ac2304}, \href {https://ui.adsabs.harvard.edu/abs/2021ApJ...922...68S} {922, 68}

\bibitem[\protect\citeauthoryear{{Shingles} et~al.,}{{Shingles} et~al.}{2020}]{Shingles2020}
{Shingles} L.~J.,  et~al., 2020, \mn@doi [\mnras] {10.1093/mnras/stz3412}, \href {https://ui.adsabs.harvard.edu/abs/2020MNRAS.492.2029S} {492, 2029}

\bibitem[\protect\citeauthoryear{{Shingles}, {Fl{\"o}rs}, {Sim}, {Collins}, {R{\"o}pke}, {Seitenzahl}  \& {Shen}}{{Shingles} et~al.}{2022}]{Shingles2022}
{Shingles} L.~J.,  {Fl{\"o}rs} A.,  {Sim} S.~A.,  {Collins} C.~E.,  {R{\"o}pke} F.~K.,  {Seitenzahl} I.~R.,   {Shen} K.~J.,  2022, \mn@doi [\mnras] {10.1093/mnras/stac902}, \href {https://ui.adsabs.harvard.edu/abs/2022MNRAS.512.6150S} {512, 6150}

\bibitem[\protect\citeauthoryear{{Shull} \& {van Steenberg}}{{Shull} \& {van Steenberg}}{1982}]{Shull1982}
{Shull} J.~M.,  {van Steenberg} M.,  1982, \mn@doi [\apjs] {10.1086/190769}, \href {https://ui.adsabs.harvard.edu/abs/1982ApJS...48...95S} {48, 95}

\bibitem[\protect\citeauthoryear{{Sim}, {R{\"o}pke}, {Hillebrandt}, {Kromer}, {Pakmor}, {Fink}, {Ruiter}  \& {Seitenzahl}}{{Sim} et~al.}{2010}]{Sim2010}
{Sim} S.~A.,  {R{\"o}pke} F.~K.,  {Hillebrandt} W.,  {Kromer} M.,  {Pakmor} R.,  {Fink} M.,  {Ruiter} A.~J.,   {Seitenzahl} I.~R.,  2010, \mn@doi [\apjl] {10.1088/2041-8205/714/1/L52}, \href {https://ui.adsabs.harvard.edu/abs/2010ApJ...714L..52S} {714, L52}

\bibitem[\protect\citeauthoryear{{Stritzinger}, {Leibundgut}, {Walch}  \& {Contardo}}{{Stritzinger} et~al.}{2006}]{Stritzinger2006}
{Stritzinger} M.,  {Leibundgut} B.,  {Walch} S.,   {Contardo} G.,  2006, \mn@doi [\aap] {10.1051/0004-6361:20053652}, \href {https://ui.adsabs.harvard.edu/abs/2006A&A...450..241S} {450, 241}

\bibitem[\protect\citeauthoryear{{Stritzinger} et~al.,}{{Stritzinger} et~al.}{2011}]{Stritzinger2011}
{Stritzinger} M.~D.,  et~al., 2011, \mn@doi [\aj] {10.1088/0004-6256/142/5/156}, \href {https://ui.adsabs.harvard.edu/abs/2011AJ....142..156S} {142, 156}

\bibitem[\protect\citeauthoryear{{Utrobin}}{{Utrobin}}{2004}]{Utrobin2004}
{Utrobin} V.~P.,  2004, \mn@doi [Astronomy Letters] {10.1134/1.1738152}, \href {https://ui.adsabs.harvard.edu/abs/2004AstL...30..293U} {30, 293}

\bibitem[\protect\citeauthoryear{{Wollaeger}, {van Rossum}, {Graziani}, {Couch}, {Jordan}, {Lamb}  \& {Moses}}{{Wollaeger} et~al.}{2013}]{Wollaeger2013}
{Wollaeger} R.~T.,  {van Rossum} D.~R.,  {Graziani} C.,  {Couch} S.~M.,  {Jordan} George~C. I.,  {Lamb} D.~Q.,   {Moses} G.~A.,  2013, \mn@doi [\apjs] {10.1088/0067-0049/209/2/36}, \href {https://ui.adsabs.harvard.edu/abs/2013ApJS..209...36W} {209, 36}

\bibitem[\protect\citeauthoryear{{Woosley}, {Heger}  \& {Weaver}}{{Woosley} et~al.}{2002}]{Woosley2002}
{Woosley} S.~E.,  {Heger} A.,   {Weaver} T.~A.,  2002, \mn@doi [Reviews of Modern Physics] {10.1103/RevModPhys.74.1015}, \href {https://ui.adsabs.harvard.edu/abs/2002RvMP...74.1015W} {74, 1015}

\bibitem[\protect\citeauthoryear{{Wygoda}, {Elbaz}  \& {Katz}}{{Wygoda} et~al.}{2019a}]{Wygoda2019}
{Wygoda} N.,  {Elbaz} Y.,   {Katz} B.,  2019a, \mn@doi [\mnras] {10.1093/mnras/stz145}, \href {https://ui.adsabs.harvard.edu/abs/2019MNRAS.484.3941W} {484, 3941}

\bibitem[\protect\citeauthoryear{{Wygoda}, {Elbaz}  \& {Katz}}{{Wygoda} et~al.}{2019b}]{Wygoda2019b}
{Wygoda} N.,  {Elbaz} Y.,   {Katz} B.,  2019b, \mn@doi [\mnras] {10.1093/mnras/stz146}, \href {https://ui.adsabs.harvard.edu/abs/2019MNRAS.484.3951W} {484, 3951}

\makeatother
\end{thebibliography}



\appendix

\section{Emergence of Non-thermal ionization}
\label{app:NT_ionization}

In this appendix, we use simple analytic arguments to estimate the epoch at which the ionization rate from non-thermal leptons becomes significant, leading to a departure from LTE for the ionization levels. The analysis presented here does not attempt to calculate NLTE effects in full, but provide rough, back-of-the-envelope calculations. For a more complete treatment of NLTE processes and their effect on SNe light curves, see, e.g., \cite{Dessart2014radiative,Shen2021NLTE,Shingles2022}.

The ionization fraction $x_i$ for a given element at ionization level $i$ is determined from the equation \citep{Jerkstrand2011PhD}
\begin{equation}
    \frac{dx_i}{dt}=\Gamma_{\text{ion},i-1}x_{i-1}+\Gamma_{\text{rec},i+1}x_{i+1}-(\Gamma_{\text{ion},i}x_{i}+\Gamma_{\text{rec},i}x_i),
\end{equation}
where $\Gamma_{\text{ion},i}$ and $\Gamma_{\text{rec},i}$ are the total ionization and recombination rates per particle from level $i$, respectively. Since the ionization and the recombination time scales are much shorter than the dynamical time, we look for the steady-state solution:
\begin{equation}
\label{eq:ion_eq1}
    \Gamma_{\text{ion},i}x_i=\Gamma_{\text{rec},i+1}x_{i+1}.
\end{equation}
At early times, when non-thermal effects are still negligible and the ionization levels are at their LTE values, the ionization rate is primarily due to photoionization. As the ejecta expands and cools, non-thermal ionization caused by the impact of fast leptons will start to dominate once its rate becomes comparable to the recombination rate. 

To estimate the time when non-thermal ionization becomes significant, we examine the state of the iron-group elements in the ejecta's core, composed of Ni, Co, and Fe. For simplicity, we consider only Fe ionization levels in what follows. We assume that the iron in the ejecta is composed only of Fe III and Fe IV, with relative abundances $x_{_\textrm{III}}$ and $x_{_\textrm{IV}}=1-x_{_\textrm{III}}$, respectively. These are the two dominant ions after 20 days from the explosion, as seen in Figure~\ref{fig:ion_compare}. The epoch $t_\text{eq}$ when the non-thermal ionization rate becomes significant is estimated by
\begin{equation}
\label{eq:ion_eq}
    \Gamma^\text{nt}_\text{ion}(t_\text{eq})x_{_\textrm{III}}(t_\text{eq})=\Gamma_\text{rec}(t_\text{eq})x_{_\textrm{IV}}(t_\text{eq}),
\end{equation}
where $\Gamma^\text{nt}_\text{ion}$ is the non-thermal ionization rate per particle for Fe III, $\Gamma_\text{rec}$ is the recombination rate per particle for Fe IV, and the ionization levels correspond to their LTE evolution within the ejecta. We now estimate the recombination and ionization rate per particle of these ions.

Assuming that recombination is primarily from free electrons within the considered temperature range \citep[]{Jerkstrand2011PhD}, the recombination rate is given by
\begin{equation}
    \Gamma_\text{rec}(t)=\alpha n_e(t),
\end{equation}
where $\alpha$ is the temperature-dependent recombination coefficient and $n_e$ is the electron density. At a temperature of $T{\sim}10^4\,\text{K}$, typical for the ejecta environment ${\approx}30$ days after explosion, $\alpha\sim1\text{--}5\times10^{-12}\,\text{cm}^3\,\text{s}^{-1}$ for doubly- or triply-ionized iron-group ions \citep{Shull1982,Nahar1996,Jerkstrand2011PhD}. The electron density is given by
\begin{equation}
    n_e=\chi_e n_\text{ion} = \chi_e\frac{1}{A}\frac{M_\text{core}}{m_p}\frac{3}{4\pi(v_\text{ej}\cdot t)^3},
\end{equation}
where $\chi_e$ is the mean number of free electrons per atom, $n_\text{ion}$ is the ion density, $A=56$ is the atomic mass number, $m_p$ is the proton mass, $M_\text{core}$ is the total mass of the core, and $v_\text{ej}$ is the characteristic velocity of the ejecta's core. Assuming the degree of ionization is similar for different elements in the ejecta, then $\chi_e=2x_{_\textrm{III}}+3x_{_\textrm{IV}}$. Unless large amounts of non-radioactive isotopes (e.g., $^{54}$Fe) are produced in the explosion, $M_\text{core}\approx\mni$, so we use $\mni$ to denote the core mass in what follows.

To estimate the non-thermal ionization rate, we assume that a fraction $\eta_\text{ion}$ of the energy of the fast leptons is converted into ionization of the ions, while the remaining energy goes into heating of the free electrons and excitation of bound electrons. The value of $\eta_\text{ion}$ depends primarily on the fraction of free electrons and the composition and only weakly on the number densities. For SNe Ia ejecta, $\eta_\text{ion}\approx0.02-0.2$ \citep{Kozma1992,Dessart2012,Li2012,Shingles2022}.
Assuming, for simplicity, that the energy for non-thermal ionization is distributed among the elements and ions in proportion to their relative abundance due to similar cross sections \citep[]{Arnaud1985}, the non-thermal ionization rate per particle will be
\begin{equation}
    \Gamma^\text{nt}_\text{ion}(t)=\eta_\text{ion}\frac{Q_\text{dep}(t)}{I_{_\textrm{III}}\cdot N_\text{ion}} = \eta_\text{ion}\frac{Q_\text{dep}(t)}{\mni}\frac{m_p A}{I_{_\textrm{III}}},
\end{equation}
where $Q_\text{dep}(t)$ is the total energy deposition of the radioactive decay, $I_{_\textrm{III}}=30.65\,\text{eV}$ is the Fe III ionization energy, $N_\text{ion}$ is the total number of ions, and it is assumed that all radioactive energy is deposited in the core. The term $\widetilde{Q}_\text{dep}(t) \equiv Q_\text{dep}(t)/\mni$ is independent of $\mni$, varying only due to differences in the $\gamma$-ray deposition fraction. For the observed range of $\gamma$-ray escape times of SNe Ia, $t_0\approx30-45$ day \citep{Sharon2020}, the range of $\widetilde{Q}_\text{dep}(30\,\textrm{day})$ varies by at most $\approx30$ percent. Plugging in the expressions for the rates in Equation~\eqref{eq:ion_eq}, and using $x_{_\textrm{III}}=1-x_{_\textrm{IV}}$, we get
 \begin{equation}
 \label{eq:t_eq}
     t_\text{eq}^3 \widetilde{Q}_\text{dep}(t_\text{eq})=\frac{3}{4\pi v_\text{ej}^3}\frac{\alpha \mni I_{_\textrm{III}}}{\eta_\text{ion}} \left(\frac{1}{A\,m_p}\right)^2\left(2+x_{_\textrm{IV}}\right) \frac{x_{_\textrm{IV}}}{1-x_{_\textrm{IV}}}.
 \end{equation}
 
We next demonstrate that Equation~\eqref{eq:t_eq} is consistent with $t_\text{eq}\approx30\,\rm{days}$ observed in Section~\ref{sec:shape_parameter} for the toy06 model from \cite{Blondin2022}. For the toy06 model, we find $\widetilde{Q}_\text{dep}(t)\approx\widetilde{Q}_\text{dep}(t=30\,\text{day})\approx1.03\times10^{43}\,\frac{\text{erg}\,\text{s}^{-1}}{M_\odot}$, with ${\approx}90$ percent of this energy deposited in the core. Using $\alpha=5.1\times10^{-12}\,\text{cm}^3\,\text{s}^{-1}$ for Fe IV recombination \citep{Nahar1996}, we get
\begin{equation}
 \begin{aligned}
     t_\text{eq}\approx& \left(\frac{0.05}{\eta_\text{ion}}\right)^{1/3} \left(\frac{\mni}{0.6\,M_\odot}\right)^{1/3} \left(\frac{0.55\times10^9\,\text{cm}\,\text{s}^{-1}}{v_\text{ej}}\right) \\
     &\times\frac{1}{0.38}\left[ \left(1+\frac{x_{_\textrm{IV}}}{2} \right)\frac{x_{_\textrm{IV}}}{1-x_{_\textrm{IV}}} \right]^{1/3}\times 31.5\,\text{day},
 \end{aligned}
 \end{equation}
where we use the core mass ($\mni=0.6\,M_\odot$) and ejecta velocity values ($v=0.55\times10^9\,\text{cm}\,\text{s}^{-1}$) of the toy06 model, and normalize for $x_{_\textrm{IV}}=0.05$ (the term in the square brackets and the value preceding it cancel out for this value), which marks the onset where Fe ionization in the toy06 model diverges from LTE (Figure~\ref{fig:ion_compare}). The good agreement with Figure~\ref{fig:ion_compare} supports our simplified description of the deviation from LTE. Note the weak dependence of $t_{\textrm{eq}}$ to the model parameters, except for $v_\textrm{ej}$ with linear dependence. 

\begin{figure}
     \centering
     \includegraphics[width=\columnwidth]{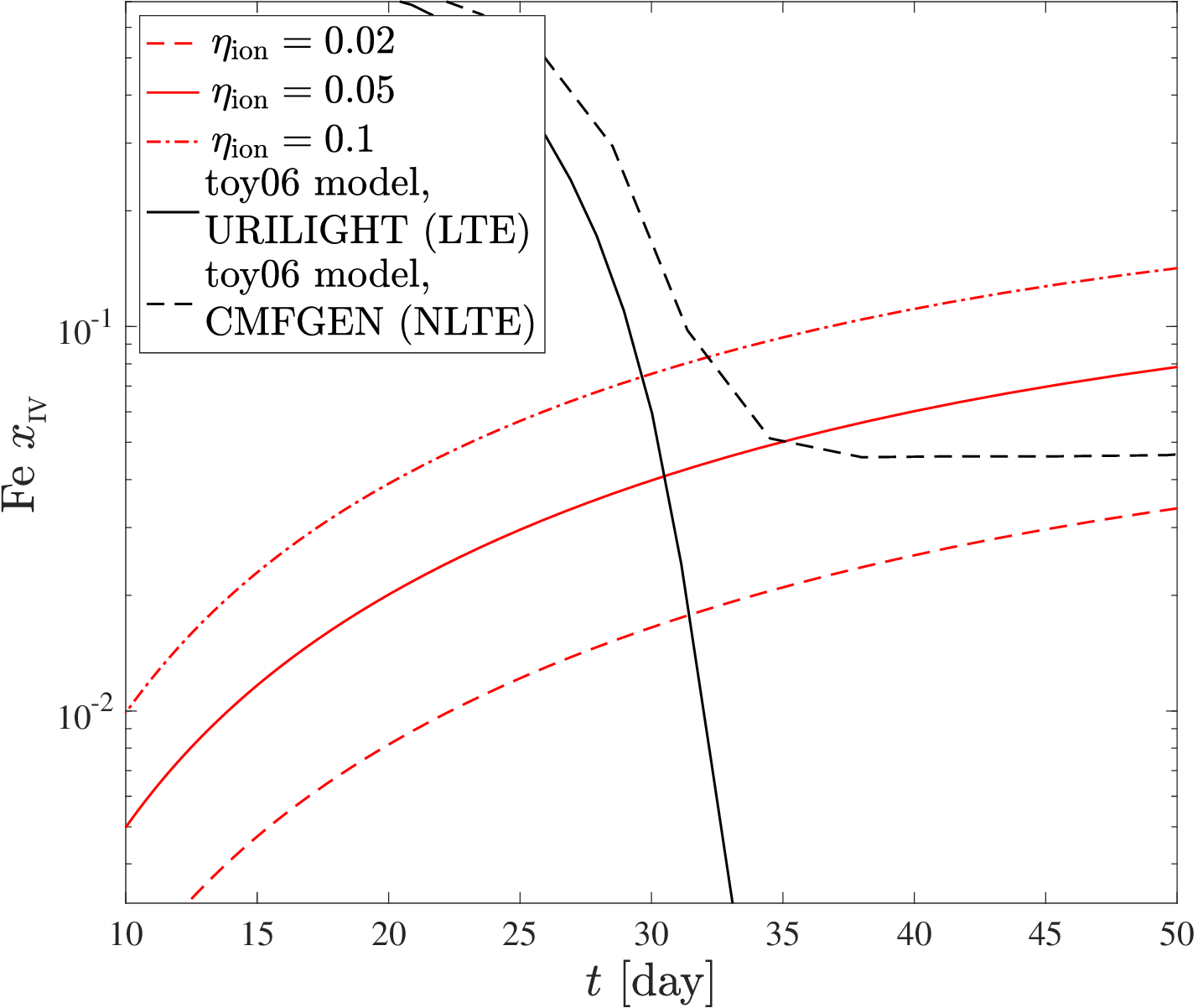}

     \caption{The evolution of the Fe IV fraction, $x_{\textrm{IV}}$, over time where non-thermal electron ionization equals the recombination rate (red lines), calculated usingEquation~\eqref{eq:t_eq} with parameters specific to the toy06 model. Various line styles represent different values of $\eta_\text{ion}$, governing the ionization fraction of non-thermal electron energy loss. The Fe IV evolution for the toy06 model simulated with URILIGHT and CMFGEN is indicated by solid and dashed black lines, respectively. The intersection between the solution of Equation~\eqref{eq:ion_eq} and the simulated evolution marks the epoch when the non-thermal ionization rate begins to dominate over the thermal ionization rate.}
     \label{fig:t_eq}
 \end{figure}

For a more accurate treatment, we present in Figure~\ref{fig:t_eq} solutions of Equation~\eqref{eq:t_eq} for several $\eta_\text{ion}$ values, shown as red lines. The \nickel\ mass, velocity, and deposition fraction are that of the toy06 model. The temperature is assumed to remain constant at $10^4\,$K, thereby maintaining the recombination coefficient constant. The solid and dashed black curves trace the evolution of $x_{_\textrm{IV}}$ in the URILIGHT and CMFGEN codes, respectively, for the toy06 model. The intersection between the solution of Equation~\eqref{eq:t_eq} and the LTE evolution of $x_{_\textrm{IV}}$ marks the epoch where non-thermal ionization equals the recombination rate, becoming non-negligible. As illustrated in Figure~\ref{fig:t_eq}, $x_{_\textrm{IV}}$ drops rapidly within a narrow time range, occurring between 25 to 35 days from the explosion. Consequently, the value of $t_\text{eq}$ at the intersection has a weak dependence on the parameters of Equation~\eqref{eq:t_eq}. Figure~\ref{fig:t_eq} also shows that these intersection times are consistent with the departure of $x_{_\textrm{IV}}$ from LTE value, as evidenced by its evolution simulated with CMFGEN. 

\begin{figure}
     \centering
     \includegraphics[width=\columnwidth]{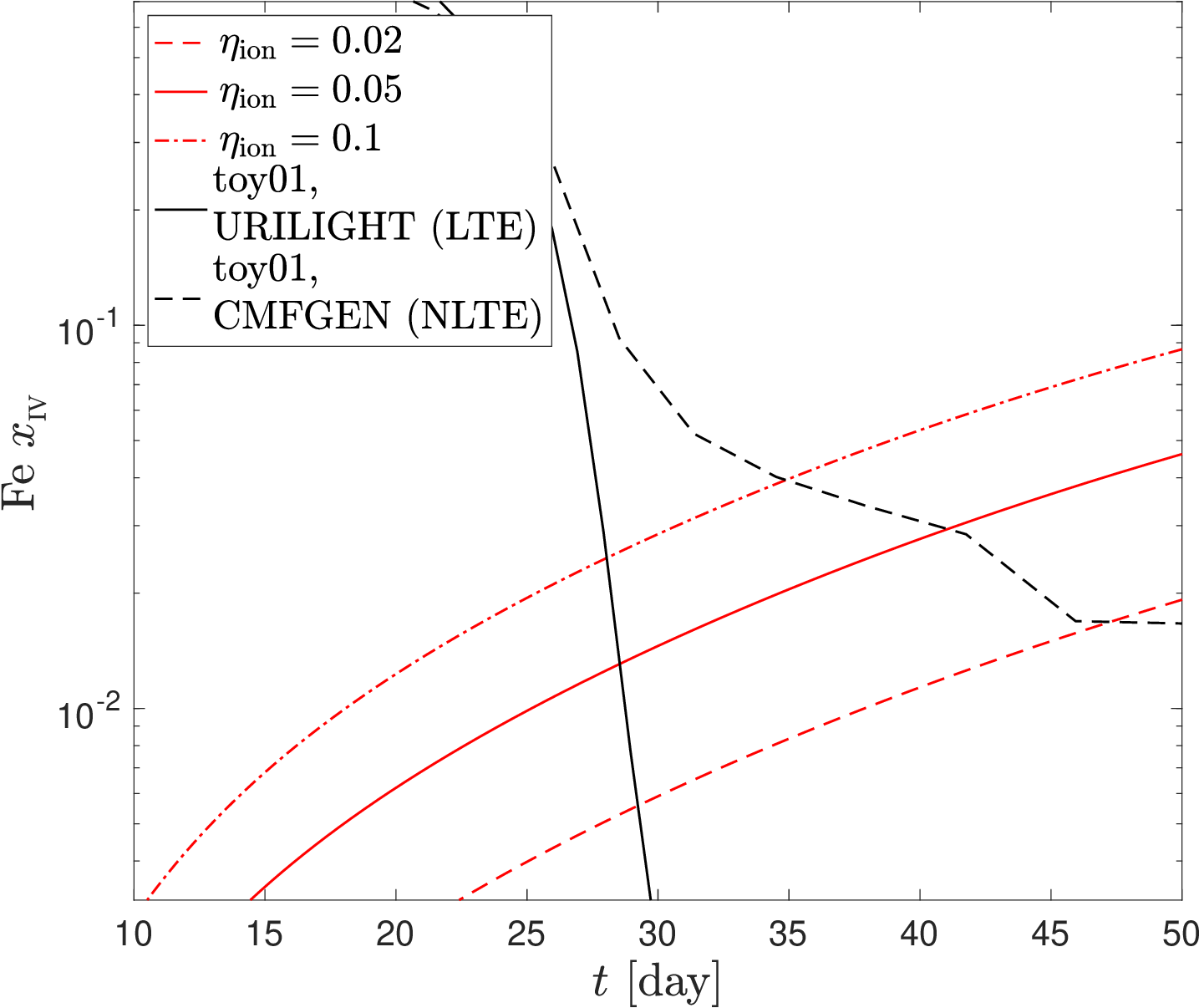}
     \caption{Same as Figure~\ref{fig:t_eq} but for the toy01 model. The results for this model are also consistent with the departure of $x_{_\textrm{IV}}$ from its LTE value occurring slightly before $30\,\rm{days}$.}
     \label{fig:t_eq_01}
 \end{figure}

For completeness, we perform the same analysis for the low-luminosity model toy01. We replace the model parameters such that the \nickel\ mass and velocity are $\mni=0.1\,M_\odot$, $v=0.24\times10^9\,\text{cm}\,\text{s}^{-1}$. In addition, due to the model's low core mass, the energy deposition $\widetilde{Q}_\text{dep}$ takes into account that a lower fraction of the energy deposition goes into the core, around ${\approx}65$ percent, than in the toy06 model. The results for the toy01 model are shown in Figure~\ref{fig:t_eq_01}, and are also consistent with the departure of $x_{_\textrm{IV}}$ from its LTE value slightly before $30\,\rm{days}$.


\section{Comparison of the rise time to analytic results}
\label{app:opacity}
In this appendix, we relate the rise time of our 1D models to the opacity of the ejecta. For this purpose, we use the analytical results of \cite[][hereafter analytical results]{KushnirKatz2019}, which provide a solution for diffusion in an isotropic, homologously expanding ejecta, under radiation-dominated pressure, uniform density and opacity, $\kappa$, and local energy deposition from radioactive decay. Following the methods in \cite{KushnirKatz2019}, the luminosity emitted from the ejecta can be calculated for a given ejecta mass $M$, outer velocity $v_o$, opacity $\kappa$, and the extent of the energy generation region $x_s$, such that
\begin{equation}
    \label{eq:xs}
    \epsilon(v,t) = \frac{Q(t)}{\frac{4\pi}{3}(x_s v_o t)^3}\times\begin{cases}
                1 & v<x_sv_o \\
                0 & \text{elsewhere}
    \end{cases},
\end{equation}
where $\epsilon$ is the energy generation rate per unit volume and $Q(t)$ is the total energy generation rate at time $t$. The resulting luminosity for an impulse of energy $E_\delta$ occurring at time $t_\delta$ is given by
\begin{equation}
    \label{eq:analytical_lum_impulse}
 \begin{aligned}
    \frac{L_\delta (t;t_\delta)}{E_\delta} = &\frac{ t_\delta}{t_\text{diff}^2}\cdot\frac{6}{x_s^3}\sum_{n=1}^\infty\left(-1\right)^{n+1}\left( \frac{\sin{(n\pi x_s)}}{n\pi}-x_s\cos{(n\pi x_s)} \right) \\
    &\quad\times e^{-\frac{(n\pi)^2}{2t_\text{diff}^2}(t^2-t_\delta^2)}\cdot u(t-t_\delta),
 \end{aligned}
 \end{equation}   
 where $t_\text{diff}^2=9\kappa M/4\pi v_o c$ and $u$ is the Heaviside step function. The luminosity for a given energy generation rate $Q(t)$ is
\begin{equation}
    \label{eq:analytical_lum}
    L(t)=\int_0^t dt' Q(t')L_\delta(t;t').
\end{equation}

We proceed by comparing the analytical results with those obtained from our radiative transfer (RT) code, hereafter referred to as numerical or simulated results. Initially, we validate the analytical outcomes by simulating ejecta with uniform opacity and density, a constant volumetric energy generation rate, and local energy deposition without $\gamma$-ray transfer. Subsequently, we conduct simulations incorporating density profiles and \nickel\ distributions specific to our models, incorporating $\gamma$-ray transfer while maintaining a uniform and constant opacity of $0.2\,\text{g}\,\text{cm}^{-2}$. To ensure consistency between the simulated and analytical models, we adjust the parameters of the analytical models —- mass, velocity, and $x_s$ -— so that the ejecta mass, kinetic energy, and \nickel\ mass remain conserved. For sub-Chandra models, $x_s$ is determined such that the mass enclosed within $x_s$ matches the \nickel\ mass, given that these models exhibit a \nickel\ distribution extending outward from the center. In contrast, for Chandra models where the central ejecta comprises non-radioactive intermediate-mass elements (IGEs) and \nickel\ becomes dominant farther from the center, a single parameter $x_s$ cannot fully characterize the \nickel\ distribution. Therefore, we sum the total mass of IGE in the central regions until the radius where \nickel\ becomes the dominant ion. We then determine the extent of the IGE in the analytical model, $x_\text{IGE}$, such that the enclosed mass within $x_\text{IGE}$ equals the IGE mass at the center of the ejecta profile. The analytical model calculation involves energy generation occurring in the shell between $x_\text{IGE}$ and $x_s$, ensuring that this shell's mass corresponds to the ejecta profile's \nickel\ mass.

\begin{figure}
     \centering
     \includegraphics[width=\columnwidth]{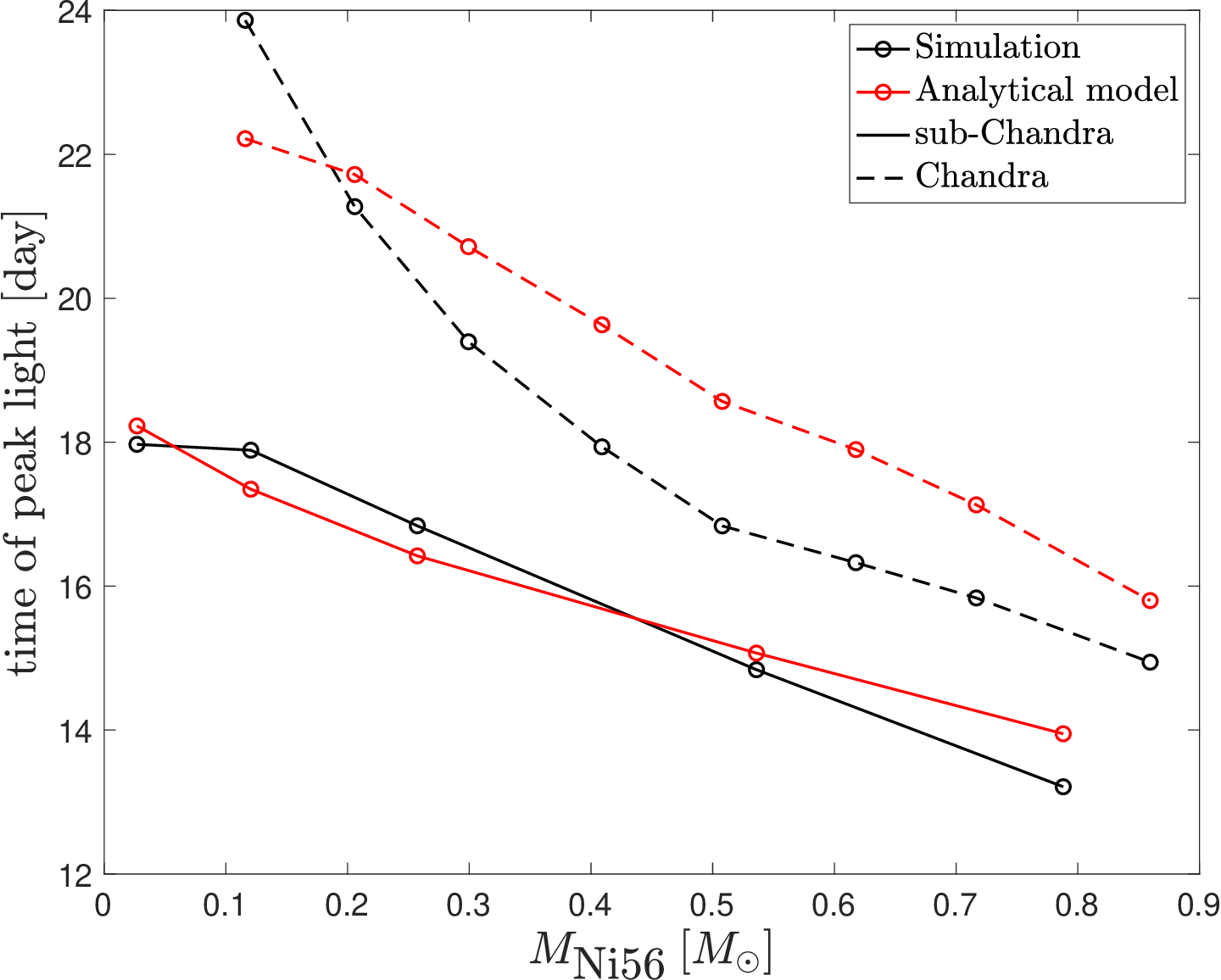}
     \caption{Epoch of peak bolometric luminosity against the synthesized \nickel\ mass, for configurations with a constant opacity of $0.2\,\text{g}\,\text{cm}^{-2}$. Black lines represent simulation results, while results from the analytic model of \protect\cite{KushnirKatz2019} are depicted in red. Solid lines correspond to sub-Chandra models, whereas dashed lines correspond to Chandra models. The parameters used in the analytical models are detailed in the text. For both models, the peak time shows a monotonic decrease with SN luminosity, consistent with the behavior of $t_0$ for these models.}
     \label{fig:tpeak_ni_const_opacity}
 \end{figure}

Figure~\ref{fig:tpeak_ni_const_opacity} shows the rise times obtained from constant-opacity simulations (black lines) and analytical models (red lines) following the prescription above. As seen in the figure, the agreement between numerical and analytical results holds within a $10$ percent margin for both sub-Chandra (solid lines) and Chandra (dashed lines) models. This robust agreement underscores the effectiveness of the analytical model. Additionally, the peak time consistently decreases with the SN luminosity across the entire range, mirroring the behavior of $t_0$ for these models. 

\begin{figure}
     \centering
     \includegraphics[width=\columnwidth]{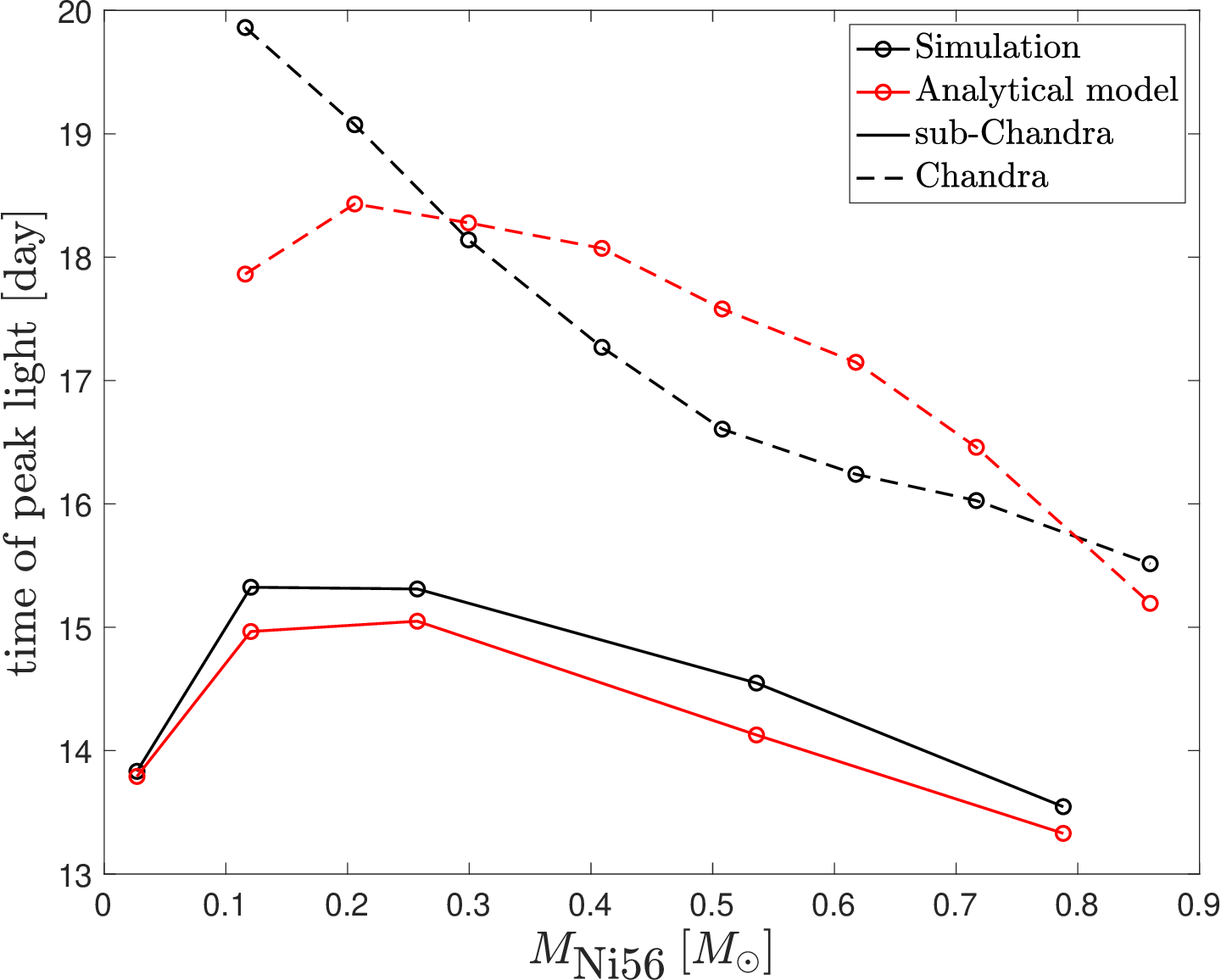}
     \caption{Same as Figure~\ref{fig:tpeak_ni_const_opacity}, but for full simulations, where opacity varies and is computed dynamically during the simulation. The opacity used in the analytical model is the mass-averaged Rosseland mean opacity derived from Equation~\eqref{eq:mean_opacity}, evaluated at peak light.}
     \label{fig:tpeak_ni}
 \end{figure}

We proceed to compare the full numerical calculations with the analytical model, accounting for varying opacity in time and space within the simulations. To align the analytical model appropriately, we employ the mass-averaged Rosseland mean opacity (Equation~\eqref{eq:mean_opacity}) at peak light, $ \left<\kappa_\textrm{R}(t_p)\right>$. The resulting opacities range from $0.1$ to $0.18\,\text{g}\,\text{cm}^{-2}$ for sub-Chandra models, and from $0.12$ to $0.18\,\text{g}\,\text{cm}^{-2}$ for Chandra models. In both models, the opacity increases monotonously with luminosity. The rise times derived from the analytical models using this approach are presented in Figure~\ref{fig:tpeak_ni}, alongside the simulation results. For sub-Chandra models, the analytical results (solid red line) closely match numerical calculations (solid black line) across the entire luminosity range, with deviations up to $\approx3$ percent. This robust agreement further underscores the influence of opacity variation on rise times and the $L_{30}/L_p$ parameter. Meanwhile, the analytical results (dashed red line) for Chandra models also align well with numerical calculations (dashed black line), albeit with deviations slightly larger, up to $\approx10$ percent.

\section{Some results in magnitude space}
\label{app:log_graphs}
\begin{figure}
    \includegraphics[width=\columnwidth]{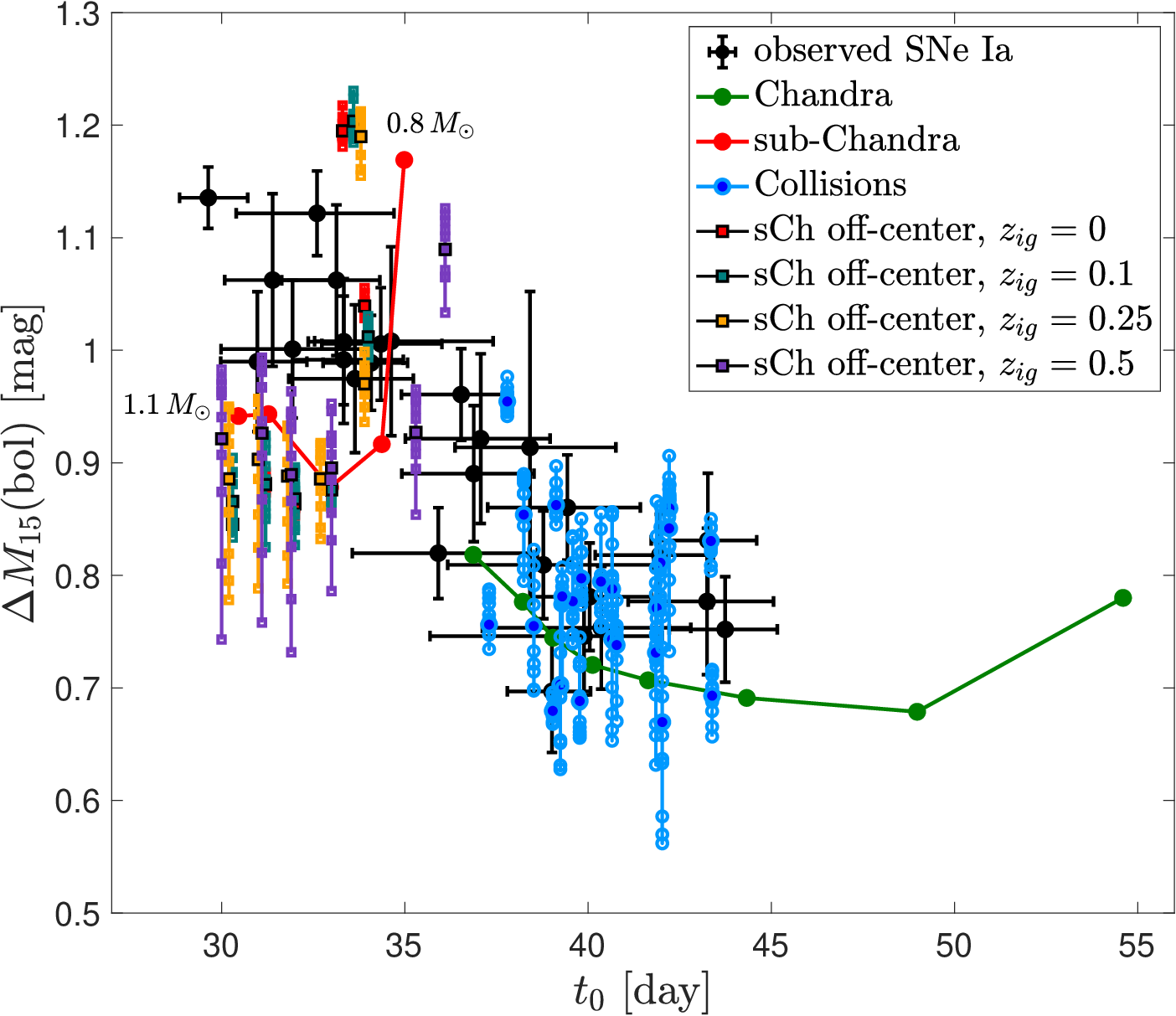}
    \caption{Same as Figure \ref{fig:t0_Lp15overLp} but $L_{p+15}/L_p$ is replaced with $\Delta M_\text{bol}(15)=-2.5\log(L_{p+15}/L_p)$.}
    \label{fig:t0_dm15}
\end{figure}

\begin{figure}
    \centering
    \includegraphics[width=\columnwidth]{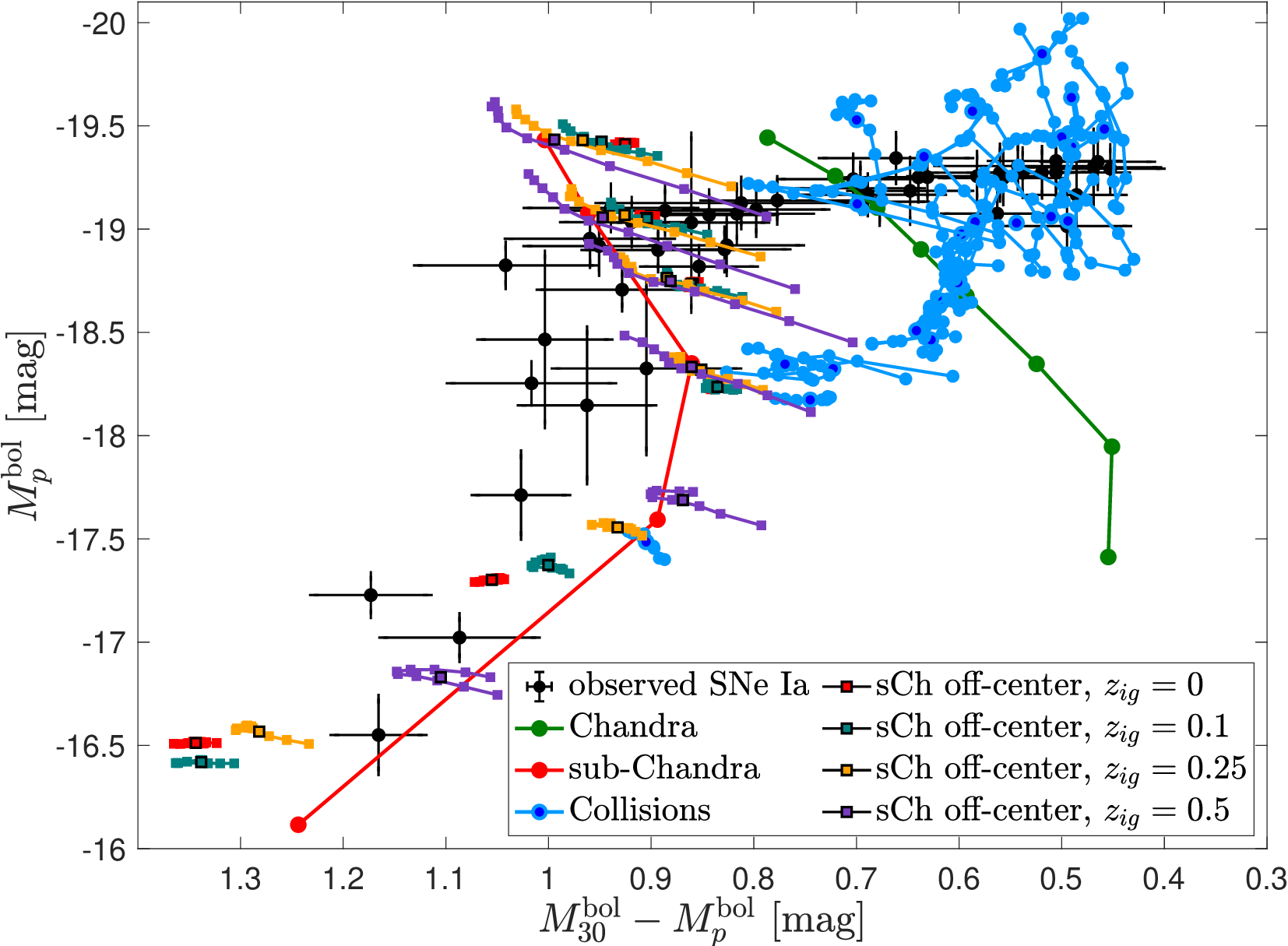}
    \caption{Same as Figure \ref{fig:Lp_LoverLp}, but in magnitude space.}
    \label{fig:Lp_LoverLp_log}
\end{figure}

Throughout this work, we used linear scaling for the luminosity and shape parameters we have presented. Here, we include additional graphs of some of our relations in magnitude space for completeness and comparison purposes. These are the $\Delta M_\text{bol}(15)$-$t_0$ relation, shown in Figure~\ref{fig:t0_dm15}, and the $L_{30}/L_p$--$L_p$ and $L_{p+15}/L_p$ relations, shown in Figures~\ref{fig:Lp_LoverLp_log} and~\ref{fig:Lp_Lp15overLp}, respectively.

\begin{figure}
    \includegraphics[width=\columnwidth]{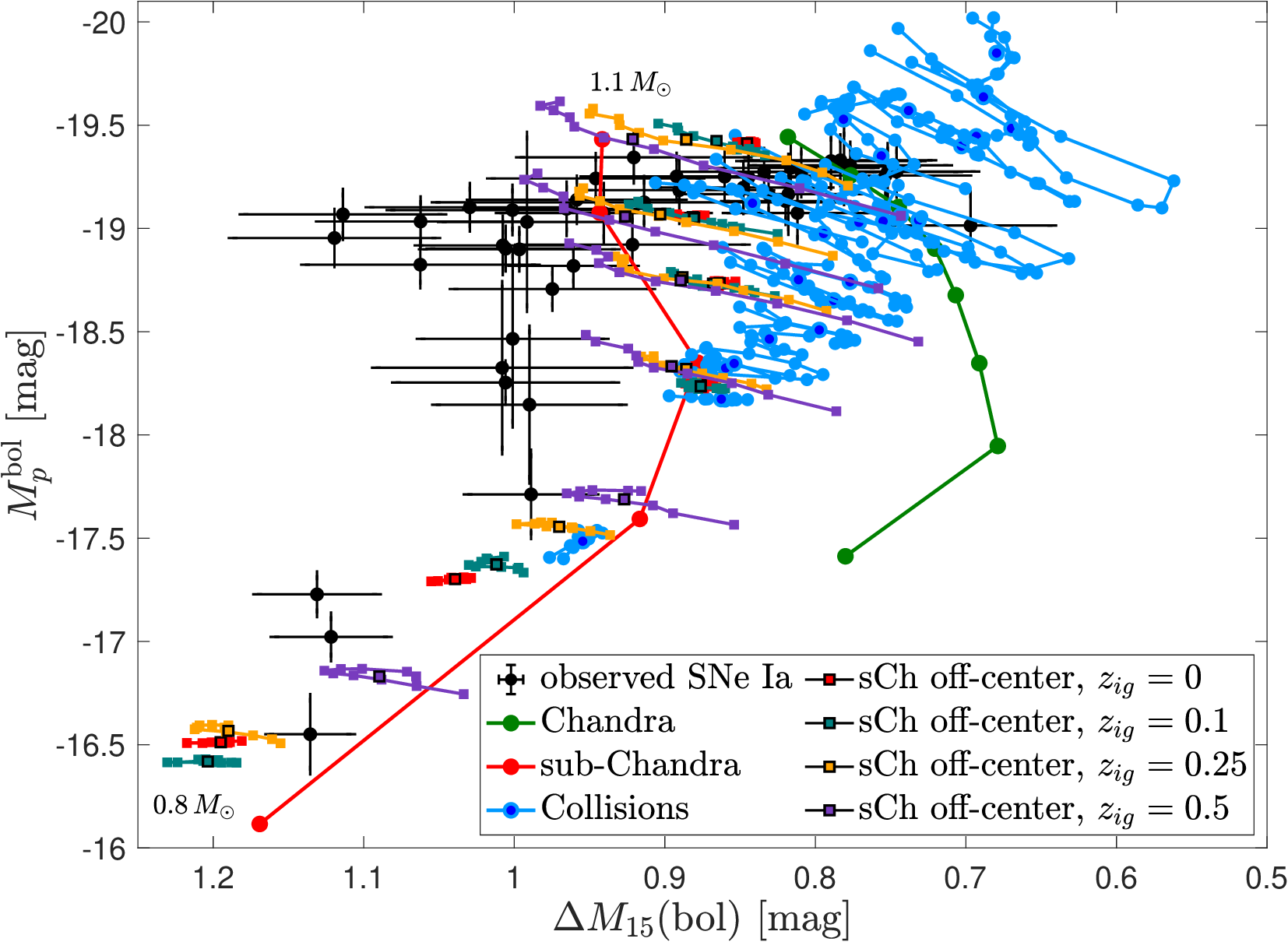}
    \caption{Same as Figure \ref{fig:Lp_Lp15overLp}, but in magnitude space.}
    \label{fig:dm_Mp}
\end{figure}
 
\section{Parameters of the SNe sample and of the models}
\label{app:parameters}

In this appendix, we outline the parameters utilized in our analysis. The characteristics of the observed sample can be found in Table~\ref{tab:sn}, while the details of the models are provided in Table~\ref{tab:models}.

\begin{table*}
 \centering
\caption{Parameters of the SNe Ia sample.}
\label{tab:sn}
 \begin{threeparttable}
	\begin{tabular}{llccccc}
 \hline 
name  & $L_p$  & $L_{30}/L_p$  & $\Delta M_{15}(\textrm{bol})$  & $M_\mathrm{Ni56}$  & $t_0$   & $t_0$ and $M_\mathrm{Ni56}$ ref.\tnote{a} \\ 
 & $(10^{43}$ $\textrm{erg}\,\text{s}^{-1})$  &   &   &  $(M_\odot)$  & (day)  &   \\ \hline 
 2003du & $ 1.36 \pm 0.12$ & $ 0.54 \pm 0.02$ & $ 0.82 \pm 0.04$ & $ 0.61_{-0.15}^{-0.19}$ & $ 35.91_{-2.70}^{-2.35}$  & 1\\ 
 2004ef & $ 1.10 \pm 0.09$ & $ 0.44 \pm 0.03$ & $ 1.00 \pm 0.06$ & $ -$ & $ -$  & $ -$\\ 
 2004eo & $ 1.12 \pm 0.13$ & $ 0.47 \pm 0.03$ & $ 0.92 \pm 0.08$ & $ 0.49_{-0.10}^{-0.12}$ & $ 37.07_{-1.89}^{-2.06}$  & 1\\ 
 2004ey & $ 1.51 \pm 0.16$ & $ 0.60 \pm 0.03$ & $ 0.84 \pm 0.05$ & $ -$ & $ -$  & $ -$\\ 
 2004gs & $ 0.92 \pm 0.07$ & $ 0.43 \pm 0.03$ & $ 0.97 \pm 0.07$ & $ 0.41_{-0.08}^{-0.08}$ & $ 33.64_{-1.59}^{-1.81}$  & 1\\ 
 2005A & $ 1.29 \pm 0.15$ & $ 0.60 \pm 0.03$ & $ 0.81 \pm 0.05$ & $ -$ & $ -$  & $ -$\\ 
 2005M & $ 1.63 \pm 0.16$ & $ 0.63 \pm 0.04$ & $ 0.78 \pm 0.06$ & $ -$ & $ -$  & $ -$\\ 
 2005cf & $ 1.43 \pm 0.19$ & $ 0.55 \pm 0.03$ & $ 0.89 \pm 0.06$ & $ 0.65_{-0.14}^{-0.13}$ & $ 36.88_{-1.64}^{-1.96}$  & 1\\ 
 2005el & $ 1.24 \pm 0.48$ & $ 0.45 \pm 0.03$ & $ 0.99 \pm 0.06$ & $ 0.49_{-0.19}^{-0.20}$ & $ 33.33_{-1.63}^{-1.40}$  & 1\\ 
 2005hc & $ 1.59 \pm 0.14$ & $ 0.61 \pm 0.04$ & $ 0.78 \pm 0.07$ & $ 0.70_{-0.10}^{-0.17}$ & $ 43.25_{-1.81}^{-2.15}$  & 2\\ 
 2005iq & $ 1.32 \pm 0.12$ & $ 0.42 \pm 0.03$ & $ 1.03 \pm 0.07$ & $ -$ & $ -$  & $ -$\\ 
 2005kc & $ 1.31 \pm 0.14$ & $ 0.48 \pm 0.03$ & $ 0.97 \pm 0.07$ & $ -$ & $ -$  & $ -$\\ 
 2005ke & $ 0.37 \pm 0.07$ & $ 0.39 \pm 0.02$ & $ 0.99 \pm 0.04$ & $ 0.13_{-0.03}^{-0.02}$ & $ 34.08_{-1.00}^{-1.31}$  & 1\\ 
 2005ki & $ 1.24 \pm 0.12$ & $ 0.43 \pm 0.03$ & $ 1.06 \pm 0.07$ & $ 0.48_{-0.06}^{-0.05}$ & $ 33.13_{-1.18}^{-1.49}$  & 1\\ 
 2006D & $ 1.11 \pm 0.13$ & $ 0.42 \pm 0.03$ & $ 1.01 \pm 0.06$ & $ 0.41_{-0.04}^{-0.06}$ & $ 33.33_{-1.40}^{-0.97}$  & 1\\ 
 2006ax & $ 1.44 \pm 0.15$ & $ 0.53 \pm 0.03$ & $ 0.85 \pm 0.06$ & $ -$ & $ -$  & $ -$\\ 
 2006bh & $ 1.15 \pm 0.13$ & $ 0.41 \pm 0.03$ & $ 1.12 \pm 0.07$ & $ -$ & $ -$  & $ -$\\ 
 2006et & $ 1.57 \pm 0.14$ & $ 0.66 \pm 0.03$ & $ 0.82 \pm 0.06$ & $ -$ & $ -$  & $ -$\\ 
 2006kf & $ 1.02 \pm 0.09$ & $ 0.38 \pm 0.03$ & $ 1.06 \pm 0.08$ & $ 0.38_{-0.04}^{-0.07}$ & $ 31.39_{-1.85}^{-1.31}$  & 1\\ 
 2006mr & $ 0.13 \pm 0.02$ & $ 0.34 \pm 0.01$ & $ 1.14 \pm 0.03$ & $ 0.04_{-0.01}^{-0.01}$ & $ 29.64_{-1.08}^{-0.78}$  & 2\\ 
 2007N & $ 0.19 \pm 0.02$ & $ 0.37 \pm 0.03$ & $ 1.12 \pm 0.04$ & $ 0.08_{-0.02}^{-0.02}$ & $ 32.61_{-2.10}^{-2.21}$  & 1\\ 
 2007S & $ 1.62 \pm 0.18$ & $ 0.65 \pm 0.03$ & $ 0.79 \pm 0.06$ & $ -$ & $ -$  & $ -$\\ 
 2007af & $ 1.02 \pm 0.08$ & $ 0.46 \pm 0.02$ & $ 0.96 \pm 0.04$ & $ 0.42_{-0.07}^{-0.06}$ & $ 36.54_{-1.21}^{-1.63}$  & 1\\ 
 2007ba & $ 0.60 \pm 0.05$ & $ 0.39 \pm 0.03$ & $ 1.01 \pm 0.07$ & $ -$ & $ -$  & $ -$\\ 
 2007bc & $ 1.31 \pm 0.13$ & $ 0.44 \pm 0.04$ & $ 1.00 \pm 0.08$ & $ -$ & $ -$  & $ -$\\ 
 2007bd & $ 1.28 \pm 0.12$ & $ 0.46 \pm 0.03$ & $ 1.11 \pm 0.07$ & $ -$ & $ -$  & $ -$\\ 
 2007le & $ 1.40 \pm 0.23$ & $ 0.64 \pm 0.03$ & $ 0.82 \pm 0.04$ & $ 0.64_{-0.12}^{-0.16}$ & $ 41.96_{-1.46}^{-1.76}$  & 1\\ 
 2007on & $ 0.55 \pm 0.18$ & $ 0.41 \pm 0.02$ & $ 0.99 \pm 0.06$ & $ 0.20_{-0.07}^{-0.08}$ & $ 30.98_{-1.35}^{-1.00}$  & 1\\ 
 2008bc & $ 1.55 \pm 0.15$ & $ 0.61 \pm 0.03$ & $ 0.81 \pm 0.05$ & $ 0.71_{-0.11}^{-0.19}$ & $ 38.78_{-3.01}^{-2.61}$  & 1\\ 
 2008bf & $ 1.52 \pm 0.15$ & $ 0.58 \pm 0.03$ & $ 0.75 \pm 0.05$ & $ 0.71_{-0.11}^{-0.07}$ & $ 40.36_{-2.44}^{-3.24}$  & 1\\ 
 2008fp & $ 1.58 \pm 0.25$ & $ 0.65 \pm 0.04$ & $ 0.78 \pm 0.05$ & $ 0.69_{-0.14}^{-0.15}$ & $ 40.04_{-1.68}^{-1.61}$  & 1\\ 
 2008gp & $ 1.50 \pm 0.15$ & $ 0.52 \pm 0.03$ & $ 0.95 \pm 0.07$ & $ -$ & $ -$  & $ -$\\ 
 2008hj & $ 1.52 \pm 0.13$ & $ 0.56 \pm 0.04$ & $ 0.89 \pm 0.06$ & $ -$ & $ -$  & $ -$\\ 
 2008hv & $ 1.10 \pm 0.09$ & $ 0.47 \pm 0.03$ & $ 1.01 \pm 0.05$ & $ 0.43_{-0.06}^{-0.07}$ & $ 34.34_{-1.67}^{-1.61}$  & 1\\ 
 2009D & $ 1.55 \pm 0.14$ & $ 0.63 \pm 0.04$ & $ 0.83 \pm 0.06$ & $ -$ & $ -$  & $ -$\\ 
 2009F & $ 0.24 \pm 0.02$ & $ 0.34 \pm 0.02$ & $ 1.13 \pm 0.04$ & $ -$ & $ -$  & $ -$\\ 
 2009Y & $ 1.55 \pm 0.17$ & $ 0.60 \pm 0.04$ & $ 0.75 \pm 0.05$ & $ 0.74_{-0.13}^{-0.12}$ & $ 43.74_{-1.42}^{-1.53}$  & 1\\ 
 2009aa & $ 1.37 \pm 0.12$ & $ 0.49 \pm 0.03$ & $ 0.96 \pm 0.07$ & $ -$ & $ -$  & $ -$\\ 
 2009ab & $ 1.29 \pm 0.15$ & $ 0.47 \pm 0.03$ & $ 0.94 \pm 0.07$ & $ -$ & $ -$  & $ -$\\ 
 2009ad & $ 1.65 \pm 0.16$ & $ 0.54 \pm 0.04$ & $ 0.92 \pm 0.08$ & $ -$ & $ -$  & $ -$\\ 
 2011fe & $ 1.22 \pm 0.17$ & $ 0.63 \pm 0.03$ & $ 0.70 \pm 0.05$ & $ 0.58_{-0.09}^{-0.09}$ & $ 39.02_{-1.05}^{-1.22}$  & 1\\ 
 2012fr & $ 1.40 \pm 0.11$ & $ 0.53 \pm 0.04$ & $ 0.83 \pm 0.06$ & $ 0.62_{-0.09}^{-0.08}$ & $ 43.27_{-1.33}^{-1.54}$  & 1\\ 
 2012ht & $ 0.65 \pm 0.24$ & $ 0.43 \pm 0.04$ & $ 1.01 \pm 0.08$ & $ 0.24_{-0.09}^{-0.10}$ & $ 34.63_{-2.78}^{-2.09}$  & 1\\ 
 2013aa & $ 1.51 \pm 0.14$ & $ 0.55 \pm 0.04$ & $ 0.86 \pm 0.05$ & $ 0.64_{-0.08}^{-0.09}$ & $ 39.44_{-1.99}^{-2.17}$  & 2\\ 
 2015F & $ 1.35 \pm 0.14$ & $ 0.47 \pm 0.06$ & $ 0.91 \pm 0.14$ & $ 0.54_{-0.10}^{-0.15}$ & $ 38.42_{-2.34}^{-2.05}$  & 1\\ 
 2015bp & $ 0.73 \pm 0.28$ & $ 0.40 \pm 0.02$ & $ 1.00 \pm 0.06$ & $ 0.29_{-0.12}^{-0.12}$ & $ 31.94_{-2.04}^{-1.95}$  & 2\\ 
 2017cbv & $ 1.55 \pm 0.16$ & $ 0.62 \pm 0.02$ & $ 0.75 \pm 0.05$ & $ 0.76_{-0.17}^{-0.10}$ & $ 39.89_{-2.13}^{-4.20}$  & 2\\ 
		\hline
	\end{tabular}
 \begin{tablenotes}
			\item [a] Reference for the values of $t_0$ and $\mni$. (1) - \cite{Sharon2020}. (2) - this work.
		\end{tablenotes}
	\end{threeparttable}
\end{table*}

\begin{table*}
 \centering
	\caption{Parameters of the models. For 2D models, the values for $L_p$ and $L_{30}/L_p$ represent angle-averaged results.}
	\label{tab:models}
	\begin{tabular}{llccccc}
 \hline 
        & name  & $L_p$  & $L_{30}/L_p$  & $\Delta M_{15}(\textrm{bol})$  & $M_\mathrm{Ni56}$  & $t_0$  \\ 
        &   & $(10^{43}$ $\textrm{erg}\,\text{s}^{-1})$  &   & (mag)  &  $(M_\odot)$  & (d)  \\ \hline 
Chandra  & DDC0\_0p5d &0.28 & 0.66 & 0.78 & 0.11 & 54.60  \\ 
        & DDC10\_0p5d &0.46 & 0.66 & 0.68 & 0.19 & 48.98  \\ 
        & DDC15\_0p5d &0.66 & 0.62 & 0.69 & 0.27 & 44.33  \\ 
        & DDC17\_0p5d &0.89 & 0.58 & 0.71 & 0.37 & 41.64  \\ 
        & DDC20\_0p5d &1.10 & 0.56 & 0.72 & 0.47 & 40.12  \\ 
        & DDC22\_0p5d &1.32 & 0.53 & 0.74 & 0.57 & 39.04  \\ 
        & DDC25\_0p5d &1.52 & 0.51 & 0.78 & 0.66 & 38.22  \\ 
        & DDC6\_0p5d &1.81 & 0.48 & 0.82 & 0.78 & 36.87  \\ \midrule
sub-Chandra,        & M085\_1Z &0.08 & 0.32 & 1.17 & 0.03 & 34.98  \\ 
central ignition        & M08\_1Z &0.33 & 0.44 & 0.92 & 0.12 & 34.37  \\ 
        & M09\_1Z &0.66 & 0.45 & 0.88 & 0.26 & 32.96  \\ 
        & M10\_1Z &1.29 & 0.41 & 0.94 & 0.54 & 31.28  \\ 
        & M11\_1Z &1.79 & 0.40 & 0.94 & 0.79 & 30.46  \\ \midrule
 sub-Chandra,       & M08\_zig0 &0.12 & 0.29 & 1.19 & 0.04 & 33.30  \\ 
 off-centered ignitions       & M08\_zig01 &0.11 & 0.29 & 1.20 & 0.03 & 33.60  \\ 
        & M08\_zig025 &0.13 & 0.31 & 1.19 & 0.04 & 33.80  \\ 
        & M08\_zig05 &0.16 & 0.36 & 1.09 & 0.06 & 36.10  \\ 
        & M085\_zig0 &0.25 & 0.38 & 1.04 & 0.09 & 33.90  \\ 
        & M085\_zig01 &0.27 & 0.40 & 1.01 & 0.10 & 34.00  \\ 
        & M085\_zig025 &0.32 & 0.42 & 0.97 & 0.12 & 33.90  \\ 
        & M085\_zig05 &0.36 & 0.45 & 0.93 & 0.14 & 35.30  \\ 
        & M09\_zig0 &0.60 & 0.46 & 0.88 & 0.24 & 33.00  \\ 
        & M09\_zig01 &0.59 & 0.46 & 0.88 & 0.24 & 33.00  \\ 
        & M09\_zig025 &0.64 & 0.46 & 0.89 & 0.25 & 32.70  \\ 
        & M09\_zig05 &0.65 & 0.45 & 0.90 & 0.26 & 33.00  \\ 
        & M095\_zig0 &0.94 & 0.45 & 0.86 & 0.39 & 32.00  \\ 
        & M095\_zig01 &0.94 & 0.45 & 0.87 & 0.39 & 32.00  \\ 
        & M095\_zig025 &0.97 & 0.44 & 0.89 & 0.40 & 31.80  \\ 
        & M095\_zig05 &0.95 & 0.44 & 0.89 & 0.39 & 31.90  \\ 
        & M1\_zig0 &1.26 & 0.43 & 0.88 & 0.54 & 31.20  \\ 
        & M1\_zig01 &1.26 & 0.43 & 0.88 & 0.54 & 31.20  \\ 
        & M1\_zig025 &1.28 & 0.43 & 0.90 & 0.54 & 31.00  \\ 
        & M1\_zig05 &1.27 & 0.42 & 0.93 & 0.53 & 31.10  \\ 
        & M11\_zig0 &1.76 & 0.43 & 0.85 & 0.80 & 30.30  \\ 
        & M11\_zig01 &1.78 & 0.42 & 0.87 & 0.80 & 30.30  \\ 
        & M11\_zig025 &1.79 & 0.41 & 0.89 & 0.80 & 30.20  \\ 
        & M11\_zig05 &1.79 & 0.40 & 0.92 & 0.79 & 30.00  \\ \midrule
Head-on collisions        & M05-M05 &0.30 & 0.43 & 0.95 & 0.11 & 37.79  \\ 
        & M055-M055 &0.56 & 0.50 & 0.86 & 0.22 & 39.13  \\ 
        & M07-M05 &0.64 & 0.51 & 0.86 & 0.26 & 42.22  \\ 
        & M06-M05 &0.66 & 0.49 & 0.85 & 0.27 & 38.25  \\ 
        & M08-M05 &0.73 & 0.56 & 0.83 & 0.29 & 43.35  \\ 
        & M06-M06 &0.76 & 0.55 & 0.80 & 0.32 & 39.82  \\ 
        & M07-M06 &0.87 & 0.57 & 0.79 & 0.38 & 40.66  \\ 
        & M08-M06 &0.96 & 0.58 & 0.81 & 0.38 & 41.97  \\ 
        & M064-M064 &0.95 & 0.57 & 0.78 & 0.41 & 39.58  \\ 
        & M08-M07 &1.18 & 0.58 & 0.79 & 0.48 & 40.35  \\ 
        & M09-M07 &1.24 & 0.61 & 0.77 & 0.51 & 41.87  \\ 
        & M09-M06 &1.25 & 0.63 & 0.73 & 0.50 & 41.85  \\ 
        & M07-M07 &1.24 & 0.58 & 0.76 & 0.56 & 38.52  \\ 
        & M09-M05 &1.35 & 0.53 & 0.84 & 0.69 & 42.21  \\ 
        & M09-M08 &1.27 & 0.62 & 0.74 & 0.54 & 40.66  \\ 
        & M08-M08 &1.66 & 0.56 & 0.76 & 0.74 & 37.30  \\ 
        & M09-M09 &1.74 & 0.64 & 0.70 & 0.78 & 39.24  \\ 
        & M10-M07 &1.81 & 0.63 & 0.69 & 0.83 & 43.39  \\ 
        & M10-M08 &1.88 & 0.66 & 0.67 & 0.81 & 42.03  \\ 
        & M10-M05 &1.96 & 0.52 & 0.78 & 0.82 & 39.29  \\ 
        & M10-M06 &2.03 & 0.58 & 0.74 & 0.88 & 40.78  \\ 
        & M10-M09 &2.16 & 0.64 & 0.69 & 1.00 & 39.77  \\ 
        & M10-M10 &2.63 & 0.62 & 0.68 & 1.25 & 39.04  \\ 
		\hline
	\end{tabular}
\end{table*}


\bsp	
\label{lastpage}
\end{document}